\documentclass[sigconf,10pt]{acmart}
\usepackage{fancyhdr}
\usepackage{balance}
\usepackage{multirow}
\usepackage{subfigure}
\usepackage{amsmath}
\usepackage{colortbl}
\usepackage{graphicx} 
\usepackage{caption}
\AtBeginDocument{%
  }

\acmYear{2023}\copyrightyear{2023}
\setcopyright{acmlicensed}
\acmConference[ACM MobiCom '23]{The 29th Annual International Conference on Mobile Computing and Networking}{October 2--6, 2023}{Madrid, Spain}
\acmBooktitle{The 29th Annual International Conference on Mobile Computing and Networking (ACM MobiCom '23), October 2--6, 2023, Madrid, Spain}
\acmPrice{15.00}
\acmDOI{10.1145/3570361.3592511}
\acmISBN{978-1-4503-9990-6/23/10}

\newcommand{\sysname}{{GPSMirror }}
\newcommand{\sysnamenospace}{{GPSMirror}}
\newcommand{\fig}{{Figure }}
\newcommand{\eq}{{Equation }}
\begin{document}

\title[GPSMirror]{\sysnamenospace: Expanding Accurate GPS Positioning to Shadowed and Indoor Regions with Backscatter}


\author{
\normalsize
Huixin Dong$^{\dagger}$, Yirong Xie$^{\dagger}$, Xianan Zhang$^{\dagger}$, Wei Wang$^{*}$$^{\dagger}$,~Xinyu Zhang$^{\ddagger}$, Jianhua He$^{\diamondsuit}$
} 

\affiliation{%
  \institution{\normalsize
  $^{\dagger}$Huazhong University of Science and Technology}
  \institution{\normalsize
  $^{\ddagger}$University of California San Diego, $^{\diamondsuit}$University of Essex}
  \country{}
  {\small
  \{huixin,xieyr997,xiananzhang,weiwangw\}@hust.edu.cn, xyzhang@ucsd.edu, j.he@essex.ac.uk
  }
}

\renewcommand{\shortauthors}{ Huixin Dong,~Yirong Xie,~Xianan Zhang,~Wei Wang,~Xinyu Zhang,~Jianhua He}
\begin{abstract}
Despite the prevalence of GPS services, they still suffer from intermittent positioning with poor accuracy in partially shadowed regions like urban canyons, flyover shadows, and factories' indoor areas.
Existing wisdom relies on hardware modifications of GPS receivers or power-hungry infrastructures requiring continuous plug-in power supply which is hard to provide in outdoor regions and some factories.
This paper fills the gap with \sysnamenospace, the first GPS-strengthening system that works for unmodified smartphones with the assistance of newly-designed GPS backscatter tags. The key enabling techniques in \sysname include: (i) a careful hardware design with microwatt-level power consumption that pushes the limit of backscatter sensitivity to re-radiate extremely weak GPS signals with enough coverage approaching the regulation limit; and (ii) a novel GPS positioning algorithm achieving meter-level accuracy in shadowed regions as well as expanding locatable regions under inadequate satellites where conventional algorithms fail. We build a prototype of the \sysname tags and conduct comprehensive experiments to evaluate them.
Our results show that a \sysname tag can provide coverage up to 27.7~m. \sysname achieves median positioning accuracy of 3.7~m indoors and 4.6~m in urban canyon environments, respectively.
\renewcommand{\thefootnote}{\fnsymbol{footnote}}
\footnotetext{$^{*}$Corresponding author.} 
\end{abstract}
\vspace{-0.6cm}

\vspace{-0.6cm}
 \vspace{-0.6cm}


\maketitle

\section{Introduction}
Global Positioning System (GPS) has long been advocated as an indispensable infrastructure for many location-based services, including navigation for the automobile~\cite{mezentsev2002vehicular,liu2018gnome}, pedestrians~\cite{mezentsev2005pedestrian}, ride-sharing~\cite{chen_bikegps_2018}, as well as asset tracking for smart factories~\cite{ahmed2020comparative}. 
GPS receivers generally perform well in clear line-of-sight (LoS) regions with a sufficient number of visible GPS satellites. 
Such an assumption does not hold in those partially shadowed regions, e.g., urban canyons, flyover shadows, and factories' indoor areas with iron sheds, where, unfortunately, navigation and tracking services are frequently needed. These services desire meter-level accuracy while positioning errors of conventional GPS in such shadowed regions often escalate to several hundred meters~\cite{chen_bikegps_2018}.
The U.S. Department of Transportation reports that vehicles traveling through urban areas of New York City frequently fail to obtain reliable street-level accuracy for navigation with GPS~\cite{IntelligenceUrbanCanyon}. 
Google posits that positioning in these shadowed regions remains ``the last great unsolved GPS problem''~\cite{greatUnsolveGPSProblem}. 

Extensive research efforts have been devoted to tackling the challenging GPS problem. 
Infrastructure-based solutions, including GPS relays~\cite{ROGERGPS,GPSrelayGRK,GPSrelayQGL1,GPSrelayRGA30-DV} and DGPS stations~\cite{morgan1995differential,GeoMaxZenith35} are available on the market.
However, such solutions typically require infrastructure support, with the dense deployment of outdoor GPS stations and/or indoor anchors, which can be expensive and labor-intensive. 
In particular, each GPS relay and DGPS product~\cite{GeoMaxZenith35} requires its own continuous power source, making them challenging to deploy in existing urban canyons and flyover areas.
Besides, laying power cables in a certain environment, such as gas factories and gas pipelines, is not allowed due to safety regulations~\cite{avoidDangerUnderGround}.

\begin{figure}[t]
	\centering
	\setlength{\abovecaptionskip}{0.cm}
	\subfigure[A \sysname tag increases \newline the number of visible satellites.]{
		\begin{minipage}[t]{0.45\linewidth}
			\centering
			\includegraphics[width=\linewidth]{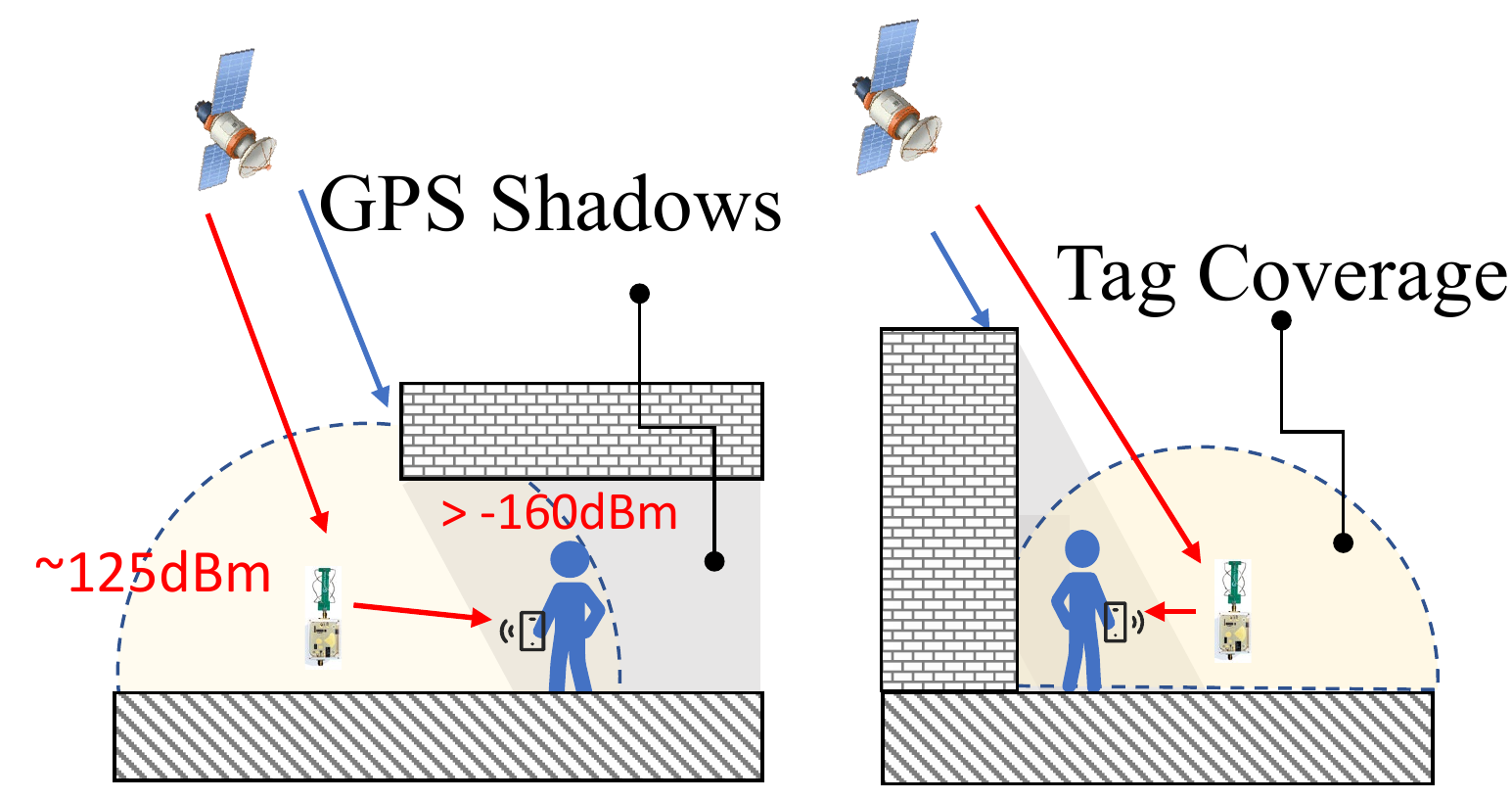}
		\end{minipage}%
	}%
	\subfigure[A~\sysname tag provides new reference propagation paths.]{
		\begin{minipage}[t]{0.46\linewidth}
			\centering
			\includegraphics[width=\linewidth]{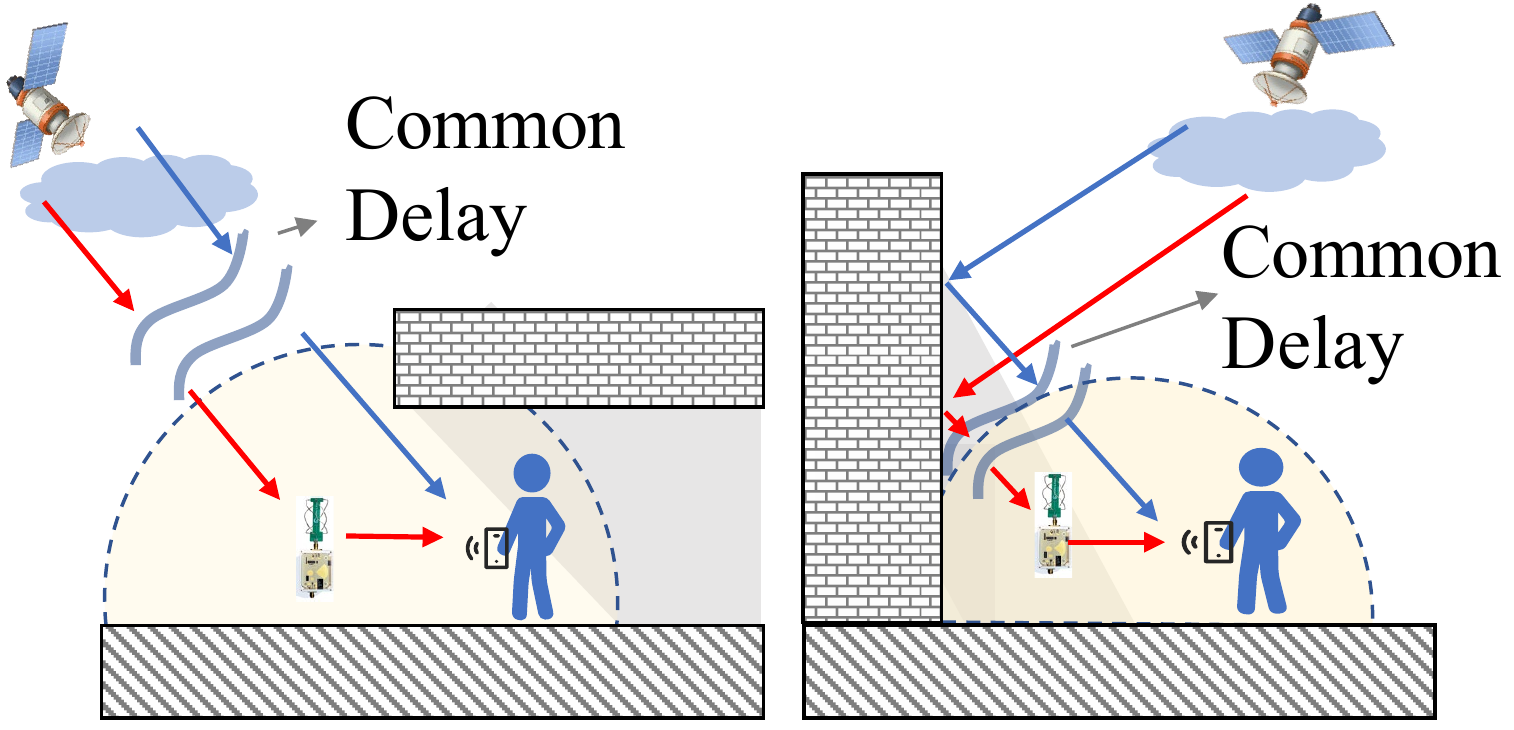}
			
		\end{minipage}%
	}%
	\caption{GPSMirror in shadowed regions.}
	\label{fig:setup}
	\vspace{-0.6cm}
\end{figure}

This paper demonstrates that it is possible to leverage passive, ultra-low-power backscatter tags to enable meter-level GPS positioning for unmodified mobile devices, e.g., smartphones, in shadowed regions. As plotted in \fig~\ref{fig:setup}, these backscatter tags can be deployed on walls, windows and unenclosed parts of flyovers to improve the GPS performance in shadowed regions. Our line of attack starts from the rationale that we can induce multipaths in a deterministic manner using backscatter, which enriches the GPS propagation features that can be exploited in positioning algorithms. Building such a system needs to fulfill requirements along the following fronts:

\textbf{(i) Sensitivity.} GPS signals are extremely weak, i.e., below -125~dBm when reaching the Earth's surface. The backscatter tags should have a high sensitivity to effectively capture and scatter such weak incidental signals.

\textbf{(ii) Coverage.} To ensure sufficient coverage, the backscatter tags need to provide sufficient gain as high as regulation permits. For example, the US regulation~\cite{NTIA_RedBook} limits the coverage of a GPS repeater's coverage within 30~m, while Europe~\cite{ETSI_Regulation} limits a GPS repeater's gain to 45~dB.

\textbf{(iii) Performance and compatibility.} The positioning algorithms must achieve comparable or higher accuracy compared to conventional GPS services, and should ideally require no hardware or firmware modification to existing GPS receivers (e.g., smartphones) to ensure wide acceptance.

Unfortunately, none of the existing systems meet the above requirements. 
On one hand, existing backscatter systems~\cite{jang2019underwater,liu2013ambient,talla2017lora} are designed to reflect strong ambient signals, e.g., those from TV towers and WiFi access points.
They cannot scatter the extremely weak GPS signals from outer space because their gain is not sufficient to overcome the fast attenuation of the weak GPS signals. While recent advances~\cite{amato2018tunnel,varshney2019tunnelscatter,varshney2020tunnel} use tunnel diodes to extend the coverage of backscatter communications, their sensitivity is still limited to around -90~dBm. 
On the other hand, existing GPS positioning algorithms~\cite{nirjon2014coin,liu2015co} are no longer applicable in our scenarios since they do not consider the extra reflection paths created by backscatter tags.

We present \sysnamenospace, the first backscatter-based system that introduces the following innovative techniques to meet the aforementioned requirements.
Inspired by recent innovations~\cite{varshney2020tunnel,amato2018tunnel}, we employ tunnel diode as a core component to build up the RF front-end of our low-power backscatter tag to achieve amplification with $\mu W$-level power consumption.
Previous designs create carrier wave~\cite{varshney2020tunnel} or modulate their bits on incident signals~\cite{varshney2019tunnelscatter} by employing a tunnel diode oscillator (TDO). However, our design target is completely different: we aim to carefully strengthen weak GPS signals according to regulations while preserving the original GPS waveforms with high fidelity to ensure correct decoding on smartphones. Instead, we propose a new design using the tunnel diode amplifier (TDA)~\cite{TunnelAmplifier}, and achieve high precision on specific gain and bandwidth control using two resonance circuits~\cite{varshney2020tunnel}. Additionally, to achieve the extreme sensitivity, \sysname hinges on a series of hardware innovations, including a circularized antenna design to capture a wide range of GPS signals while reflecting them to the shadowed regions, a coupling circuit with an open loop ring (OLR) to reduce the return loss caused by the solder of circuit component in the transmission line (TL), a noise suppression design to minimize SNR degradation caused by tunnel-diode based reflective amplifier.

We design a novel positioning algorithm that takes advantage of the tag-manipulated paths to (i) achieve positioning with comparable precision to LoS GPS areas,
and (ii) dramatically improve positioning precision to meter level in cases where conventional algorithms are significantly compromised (tens to hundreds of meters precision) in NLoS conditions. 
\sysname only requires the assistance of a single tag and works on smartphones as a software update without any hardware/firmware modification.

We prototype the \sysname tag with off-the-shelf components and benchmark its hardware performance through microscopic measurements. We further deploy \sysname tags in four real-world scenarios to verify their effectiveness. Experimental results show that a \sysname tag achieves a flat 22~dB gain in the L1 band and covers an area with a radius of about 27.7~m for smartphone reception in an urban canyon and 30~m in a corridor. The positioning algorithm achieves a median position error as low as 3.7~m in indoor scenarios and 4.6~m in an urban canyon. The main contributions are summarized as follows:

(i) \sysname is the first backscatter system that scatters signals from orbiting satellites. The hardware design pushes the limit of backscatter in that it can re-radiate weak GPS signals as low as -125~dBm with sufficient coverage.

(ii) We propose novel GPS positioning algorithms to extend GPS positioning services in shadowed or indoor regions where conventional GPS positioning fails or is severely undermined. It works seamlessly with off-the-shelf smartphones without any hardware or firmware modification.

(iii) We implement and evaluate the \sysname in real-world environments using off-the-shelf smartphones from different manufacturers. Our experiments verify the feasibility and compatibility of \sysname.


\vspace{-0.4cm}
\section{Overview}
 
The \sysname system comprises the hardware design of backscatter tags and the positioning algorithms on smartphones to fully use the signals re-radiated by the tags. 

\textbf{(i) Hardware design}. We employ two key innovations to build up the \sysname tags: a high-sensitive RF front-end and a tunnel diode-based reflection amplifier. 

\textbf{(ii) Positioning algorithm}. We present an algorithm to subtly use \sysname tags to improve the GPS positioning performance in shadowed regions. When the number of visible satellites is inadequate for conventional GPS positioning (\fig~\ref{fig:setup}~(a)), we construct virtual satellites with a single \sysname tag to guarantee coverage. When the number of visible satellites is adequate (\fig~\ref{fig:setup}~(b)), we take advantage of the valid propagation path through \sysname tags to eliminate propagation errors of non-line-of-sight (NLoS) and improve accuracy.

\begin{figure}[t]
	\centering
	\includegraphics[width=0.8\linewidth]{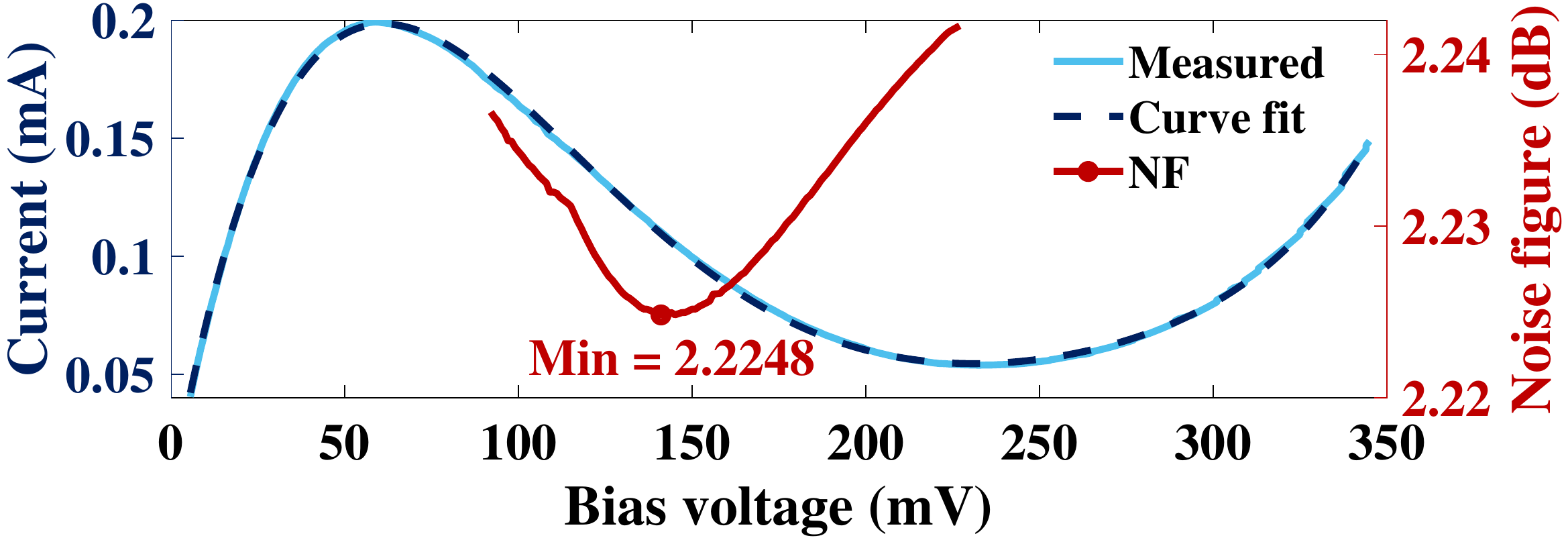}
	\caption{IV-Curve of Tunnel Diode MBD1057E28. Measured with a DC power supply (model IPS900B) and a current meter (model VC890D).}
	\label{fig:IV-curve}
  \vspace{-0.5cm}
\end{figure}

\section{Hardware Design} 
\label{sec:hardware}

\begin{figure}[t]
	\centering
	\setlength{\abovecaptionskip}{0.cm}
	\includegraphics[width=0.8\linewidth]{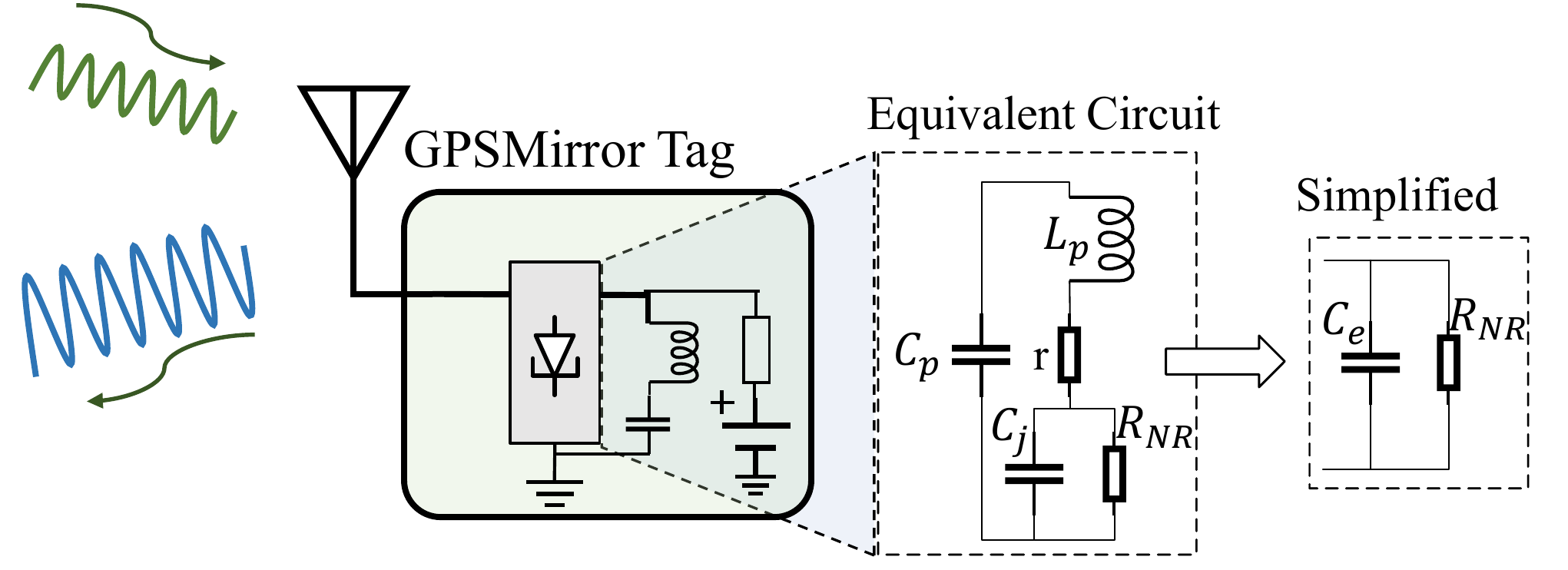}
	\caption{\sysname tag. \sysname tag enhances the injected wireless signals using a tunnel diode, which can be regarded as a combination of a negative resistance $R_{NR}$, parasitic capacitors and inductors.}
	\label{fig:model}
 \vspace{-0.6cm}
\end{figure}

This section describes the hardware design of \sysname tags that can scatter weak GPS signals with sufficient coverage.

\vspace{-0.3cm}
\subsection{Pushing Scattering Coverage to the Limit}\label{sec:Hardware_Coverage}
A fundamental challenge facing \sysname is the dramatic attenuation when backscatter tags re-radiate GPS signals, which severely limits the coverage of the \sysname tags. 
To maximize the coverage, the tags should be able to enhance the energy of the scattered GPS signals under the regulation permit~\cite{NTIA_RedBook,ETSI_Regulation}. 
Such a capability can be characterized by the standard \textit{backscatter reflection coefficient} $\Gamma$~\cite{liu2014enablingFullduplexBks,jang2019underwater}, i.e., the ratio of the amplitude of the reflected wave to the incident wave:
\begin{small}
	\begin{equation} \label{equation: reflecoefficient}
		\Gamma= \cfrac{Z_L - Z_A^{*}}{Z_L + Z_A} = \cfrac{1 - Z_A^{*}/Z_L}{1 + Z_A/Z_L}
	\end{equation}
\end{small}
where $Z_L$ and $Z_A$ stand for the impedance of the internal circuit and the antenna, respectively. 
The \textit{gain} of the backscatter tag is defined as $|\Gamma|^2$~\cite{TunnelAmplifier}. Obviously, when $Z_L$ is negative, the gain is greater than 1, i.e., the signals are amplified after re-radiated by backscatter tags.  

To achieve this, we adopt tunnel diodes as a core component, as they can realize negative resistance with $\mu W$-level power consumption. Plotted in \fig~\ref{fig:IV-curve} is the IV-Curve of the diode, which holds a region of negative resistance (NR) due to the quantum tunneling effect~\cite{varshney2019tunnelscatter}. The negative resistance ($R_{NR}$) can be calculated from the slope of the IV-Curve. For MBD1057, which is used in \sysnamenospace, $R_{NR}$ is about $-650 \Omega$ in our design with a bias voltage set around $140 mV$.

\begin{figure}[t]
	\centering
	\setlength{\abovecaptionskip}{0.cm}
        \setlength{\belowcaptionskip}{0.cm}
	\subfigure[Reflection amplifier circuit.]{
		\begin{minipage}[t]{0.5\linewidth}
			\centering
			\includegraphics[width=\linewidth]{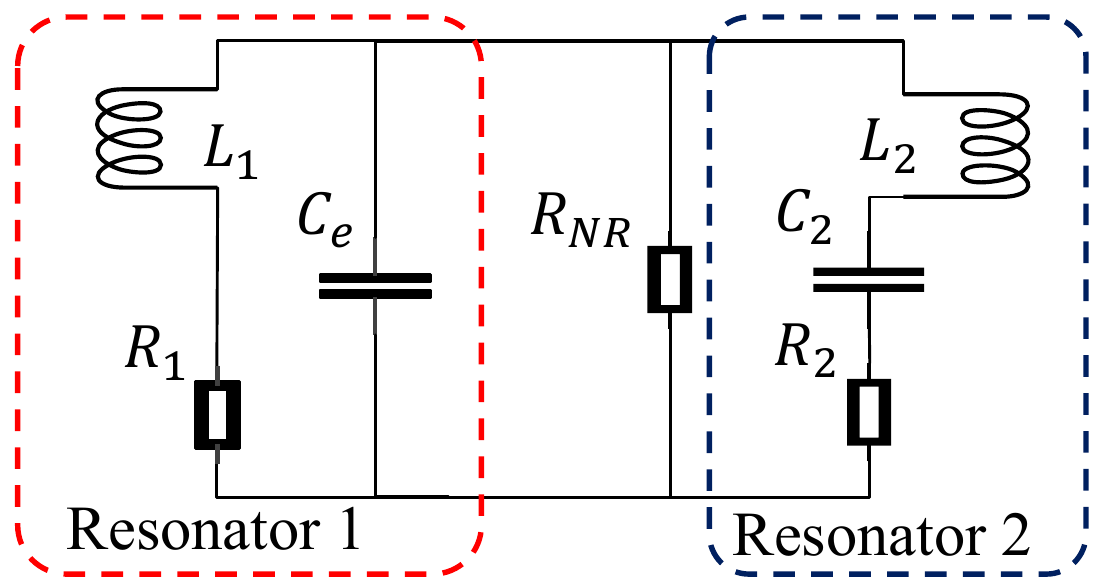}
		\end{minipage}%
	}%
	\subfigure[Bandwidth control.]{
		\begin{minipage}[t]{0.46\linewidth}
			\centering
			\includegraphics[width=\linewidth]{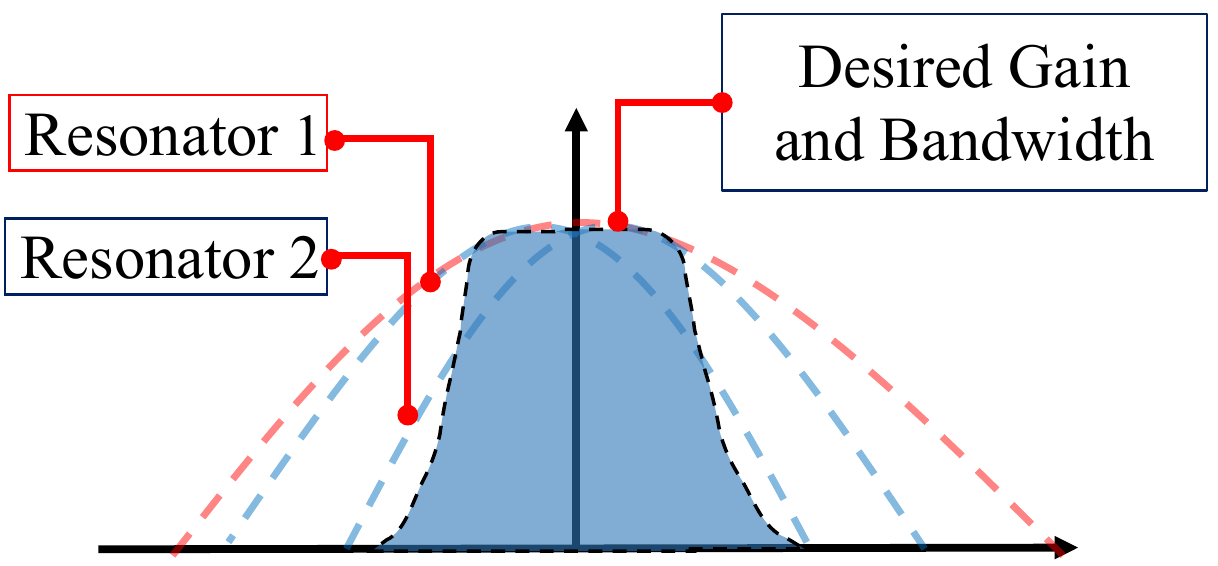}
		\end{minipage}%
	}%
	\caption{Reflection Amplifier. Two resonant circuits are used to match the impedance of the circuit and achieve the desired bandwidth.}
	\label{fig:CircuitDesign}
 \vspace{-0.7cm}
\end{figure}

Classical tunnel diode was invented in the 1950s~\cite{jonscher1961physicsTunnelInventor} and revived in 
recent backscatter systems~\cite{amato2018tunnel,varshney2020tunnel,varshney2019tunnelscatter} as a low-power amplifier.  
These systems aim to use a tunnel diode oscillator to re-modulate the incident signals, e.g., in the form of low-rate ASK or FSK, which can be decoded by a specialized demodulator. In other words, the backscatter tag's modulation differs from the signal source.  
In contrast, \sysname needs to provide a high gain while preserving the incident signal's waveform. Therefore, we employ the tunnel-diode-based reflection amplifier (TDA)~\cite{farzami2017ultra,TunnelAmplifier} to build \sysname instead.

\textbf{Gain and bandwidth.} The $R_{NR}$ of the tunnel diode is about $-650~\Omega$ while standard antennas' impedance $Z_A$ is $50~\Omega$.
To realize the amplification effect according to \eq~\ref{equation: reflecoefficient}, we need an impedance matching network to transform the impedance of the internal circuit's $Z_L$ to make it approach $-Z_A$.

Yet, the tunnel diode cannot simply be regarded as a negative resistance alone. Instead, it is equivalent to a combination of multiple parasitic elements, including internal junction capacitance $C_j$, packaging parasitic elements $C_p$, $L_p$ and $r$, whose typical values are $0.1 pF$, $0.3 pF$, $1.2 nH$ and $6 \Omega$, respectively, according to the datasheet~\cite{MBD1057Datasheet}. With the center frequency $f_c$ fixed in GPS L1 band, we simplify the 
tunnel diode circuit by series-parallel conversion step by step and then acquire an equivalent circuit as plotted in \fig~\ref{fig:model}, where the equivalent capacitor is about $0.465$~pF.

The conventional approach directly computes the overall impedance of the tunnel diode and builds a matching network. However, as the tunnel diode amplifier is susceptible to impedance changes, a straightforward search for values as in the conventional approach may introduce more parasitic parameters during manufacturing, resulting in unexpected performance. 

To address this issue, \sysname achieves desired bandwidth and gain by using \textit{two resonance circuits} instead, thereby making the circuit highly repeatable and easily debugged. The first resonant circuit aims to eliminate the effects of the parasitic elements inside the tunnel diode, while the second considers the impedance changes caused by the first resonant circuit and tunes the circuit to desired bandwidth and gain.
In a typical series ``LC'' resonator circuit, the relationship of inductance $L$, capacitance $C$ and frequency $f_c$ can be described as~\cite{zhao_nfc_2020}:
\begin{small}
	\begin{equation}
             f_c=1/2\pi\sqrt{LC}
	\end{equation}
\end{small}
Given the equivalent capacitance $C_e=0.465 pF$ inside the tunnel diode, the only suitable inductance $L_1 = 2.195 nH$ can be derived.  

Then, the gain and bandwidth can be estimated through the quality factor $Q$, i.e., the ratio of resonant frequency to bandwidth, given by
\begin{small}
	\begin{equation}
            Q=2\pi f_c L/R
	\end{equation}
\end{small}
where $R$ is the parasitic resistance inside the inductor and transmission lines. A smaller $R$ contributes to a higher $Q$, and thus we use low-loss microstrip lines to route the circuit.

The second ``RLC'' circuit aims to achieve the desired gain and bandwidth. Given the needed bandwidth and impedance, we can determine a set of inductance and capacitance to achieve our goal according to the above equations. To ensure the gain's flatness over the band, we slightly staggered the center frequency of the two resonant networks to ensure a consistent gain over the GPS band and sharp roll-off, which is the gain's decreasing speed~\cite{li2009compactRollOff}, beyond the band, as plotted in \fig~\ref{fig:CircuitDesign}~(b).
\begin{figure}[t]
	\centering
	\setlength{\abovecaptionskip}{0.cm}
	\setlength{\belowcaptionskip}{0.cm}
	\subfigure[Microstrip circuit.]{
		\begin{minipage}[t]{0.3\linewidth}
			\centering
			\includegraphics[width=\linewidth]{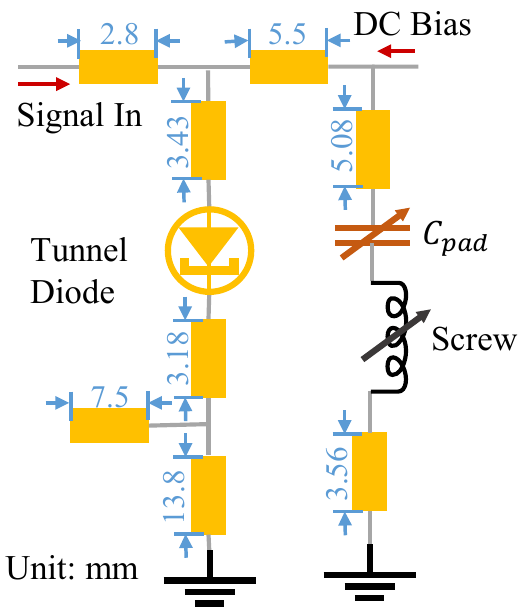}
		\end{minipage}%
	}%
	\subfigure[Stackup]{
		\begin{minipage}[t]{0.55\linewidth}
			\centering
			\includegraphics[width=\linewidth]{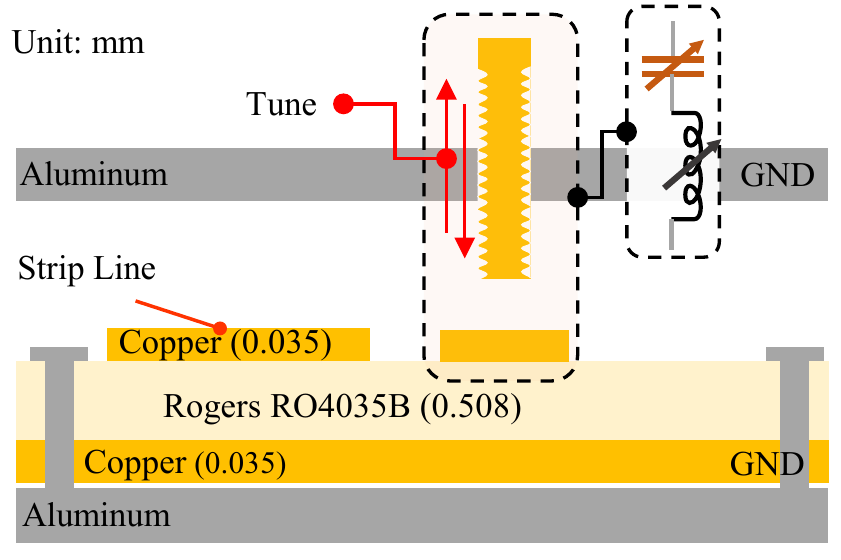}
		\end{minipage}%
	}%
	\caption{Structure of reflection amplifier circuit.}
	\label{fig:MicroStripStackup}\vspace{-0.3cm}
\end{figure}

Unfortunately, integrated inductors and capacitors contain many parasitic elements along with soldering artifacts, making it difficult for circuit diagnosis. Thus, we transform the reflection amplifier circuit, as illustrated in \fig~\ref{fig:CircuitDesign} (a), to practical microstrip lines on a PCB substrate to ensure the accuracy, as illustrated in \fig~\ref{fig:MicroStripStackup} (a).
Furthermore, to deal with the PCB manufacturing imprecision that causes the resonant frequency to deviate from the design, we employ a tuning stub on board to adjust the center frequency of the first resonator and a tune pad with a tuning screw for the second one. \fig~\ref{fig:MicroStripStackup}~(b) shows the stack-up of the 
\sysname reflection amplifier circuit. 
As the screw is closer to the tuning pad, the values of $L_2$ and $C_2$ increase and the resonant frequency of resonator 2 decreases, and vice versa.

\begin{figure}[t]
	\centering
	\setlength{\abovecaptionskip}{0.cm}
	\subfigure[GPS signals reception via different antennas.]{
		\begin{minipage}[t]{0.65\linewidth}
			\centering
			\includegraphics[width=\linewidth]{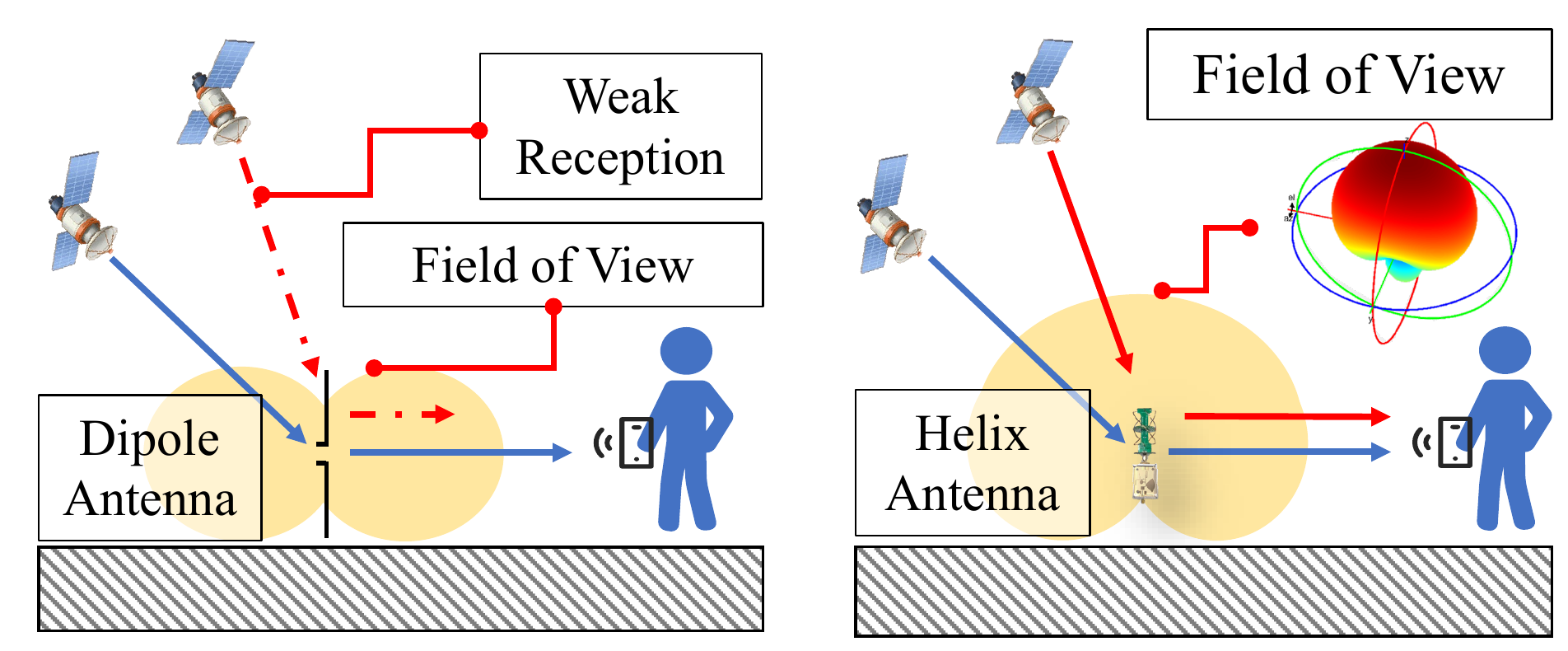}
		\end{minipage}%
	}%
	\subfigure[Field of View.]{
		\begin{minipage}[t]{0.28\linewidth}
			\centering
			\includegraphics[width=\linewidth]{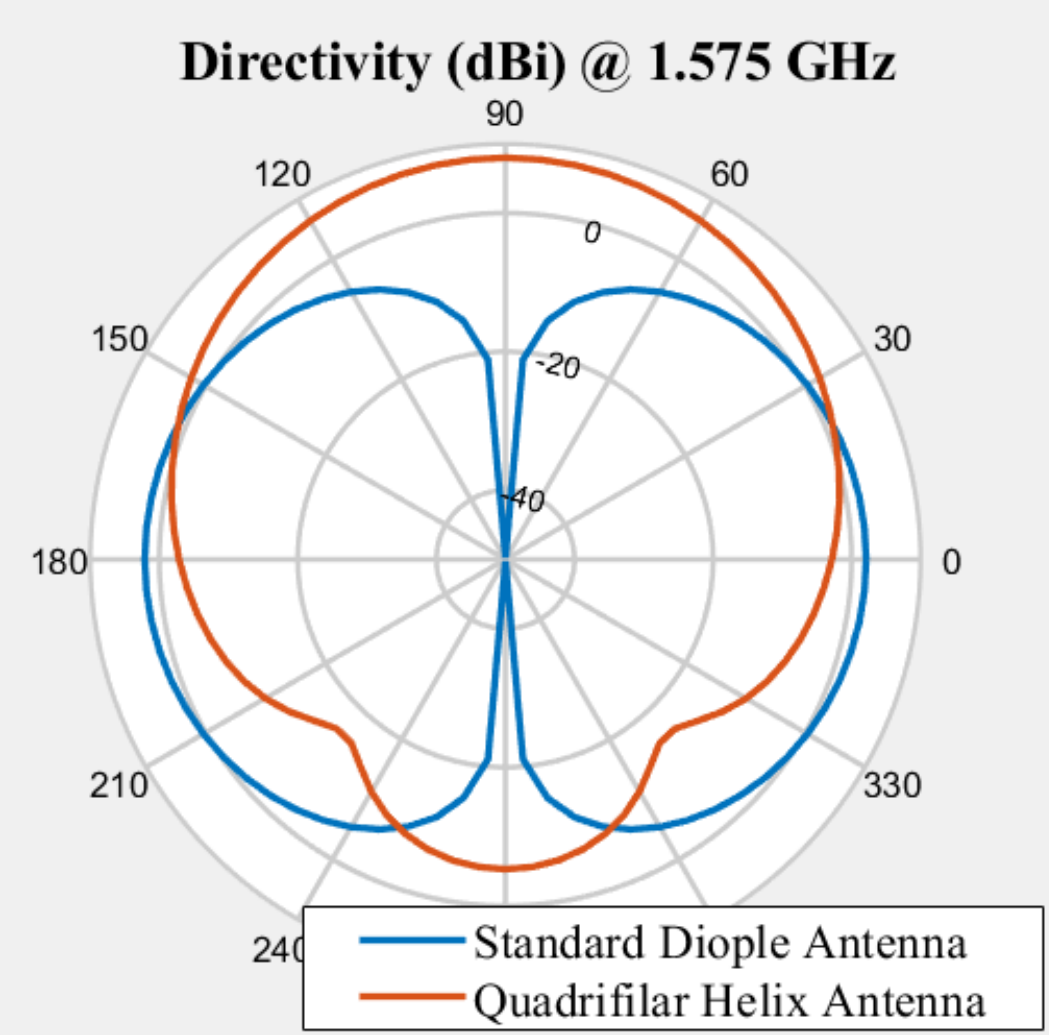}
		\end{minipage}%
	}%
	\caption{\sysname employs QHA for better FoV.}
	\label{fig:antennaComparison}\vspace{-0.6cm}
\end{figure}

\begin{figure}
	\centering
	\setlength{\abovecaptionskip}{0cm}
	\setlength{\belowcaptionskip}{0cm}
	\subfigure[Direct AC coupling]{
		\begin{minipage}[t]{0.33\linewidth}
			\centering
			\includegraphics[width=\linewidth]{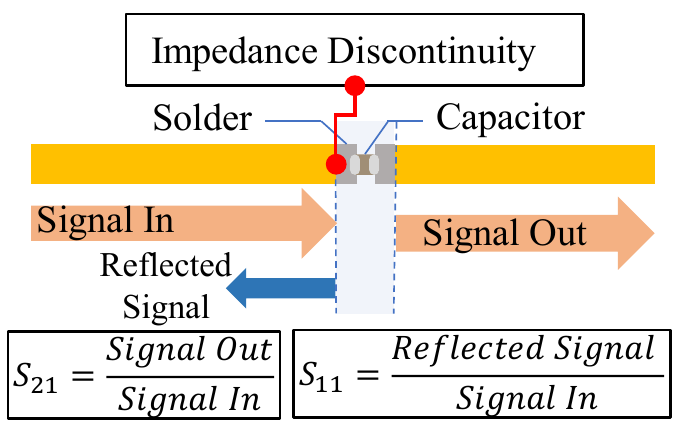}
		\end{minipage}%
	}%
	\subfigure[OLR coupling circuit]{
		\begin{minipage}[t]{0.33\linewidth}
			\centering
			\includegraphics[width=\linewidth]{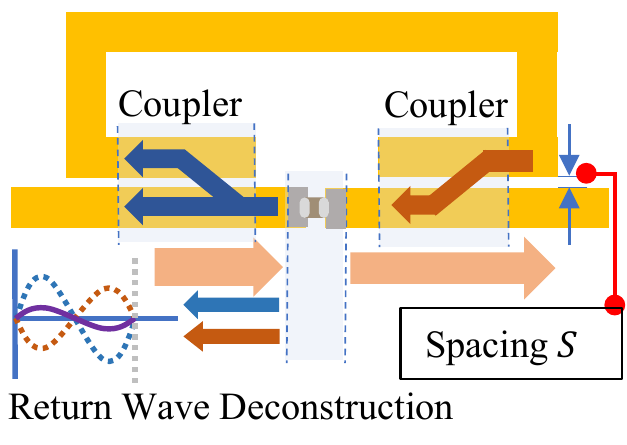}
		\end{minipage}%
	}%
	\subfigure[Loss reduction]{
		\begin{minipage}[t]{0.28\linewidth}
			\centering
			\includegraphics[width=\linewidth]{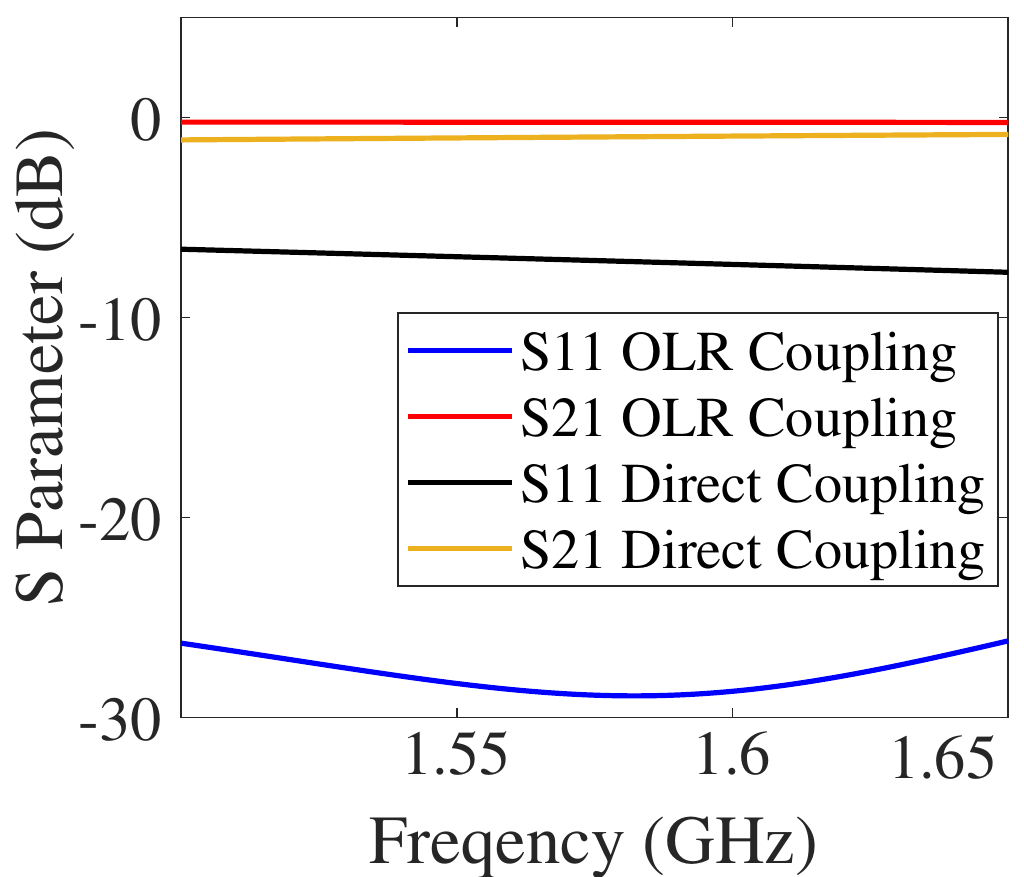}
		\end{minipage}%
	}%
	\caption{\sysname designs an open loop coupling circuit to reduce return loss in the transmission line.}
	\label{fig:couplingCircuit} \vspace{-0.6cm}
\end{figure}
\vspace{-0.3cm}
\subsection {High-Sensitivity RF Front-end for \sysname} \label{subsec:Hardware_Sensitivity}
Sensitivity~\cite{AmplifierSensitivity} is a measure of the magnitude of input signal needed for the amplifier to produce full output. 
To scatter GPS signals with sufficient gain, \sysname must be sensitive enough to acquire the weak GPS signals in the first place. 

This is different from traditional passive backscatter tags which do not need to detect and amplify the incident signals and hence do not need to optimize sensitivity. It also differs from traditional radio receivers which consider the modulation/encoding gain as part of the sensitivity. In effect, the sensitivity of \sysname is the detector sensitivity which describes how efficiently the radiation is converted into a usable signal.  
We design a highly sensitive RF front-end for \sysname via three steps, including an antenna design that maximizes the incident signals, a low loss coupling design that reduces the return loss when the signal is coupled into the internal circuit, and a low noise amplifier design that minimizes the current noise in the circuit.

\textbf{Capturing and radiating signals with a wide field-of-view (FoV).}
Traditional backscatters~\cite{liu2013ambient,zhao2019ofdma,varshney2019tunnelscatter} employ dipole antennas to scatter signals within the azimuth plane. 
In contrast, \sysname reflects GPS signals over a wide 3D FoV by adapting the classical Quadrifilar Helix Antenna (QHA) structure~\cite{4751699QHAinventor}. 

We build the antenna with a diameter of 3.8~cm, a comparable size to \sysname tags. The wavelength of the GPS signal in the L1 band is known and the remaining parameters of the antenna can be acquired from the helix antenna structure equation~\cite{4751699QHAinventor}.
We also conduct simulations in ANSYS HFSS to verify the above design choices. As plotted in \fig~\ref{fig:antennaComparison} (b), QHA significantly outperforms the standard dipole antenna in the horizontal upward direction with an average of 7.3~dB gain in all upward directions.

\textbf{Reducing return loss in the transmission line.} 
After the signals are captured by the antenna, they need to be passed to the RF front-end which may induce additional loss. 
A coupling circuit can be used to minimize such loss. Traditional backscatter tags use inductors and capacitors to build the matching network to couple the signals from the antennas to the internal circuit as shown in the \fig~\ref{fig:antennaComparison}. However, these components need to be soldered on, which tends to damage the integrity of the transmission line (TL), resulting in the return loss, i.e., signal reflection and energy dissipation before coupling into the internal circuit.

We design a new type of microstrip coupling circuit to minimize the loss of TL, as illustrated in \fig~\ref{fig:couplingCircuit}(b). We use a microstrip couple filter, which consists of two parallel TLs, to couple the returned signal to the open-loop-ring (OLR) circuit for additional transmission delays. Intuitively, when the signal coupled to the OLR circuit returns to the pad with the opposite phase, the return signal is canceled and loss minimized.    
The overall length of the loop determines the phase delay, an additional $\lambda/2$ delay contributes to maximum signal deconstruction.
We conduct an RF circuit simulation to verify our design.
The simulation results in \fig~\ref{fig:couplingCircuit}~(c) show that our design significantly reduces the return loss. The S11 parameter, which is defined in \fig~\ref{fig:couplingCircuit}~(a), is considerably lower when compared to directly soldering a capacitor.
\begin{figure}
	\centering
	\setlength{\abovecaptionskip}{0.cm}
	\setlength{\belowcaptionskip}{0.cm}
	\subfigure[\sysname schematic]{
		\begin{minipage}[t]{0.65\linewidth}
			\centering
			\includegraphics[width=\linewidth]{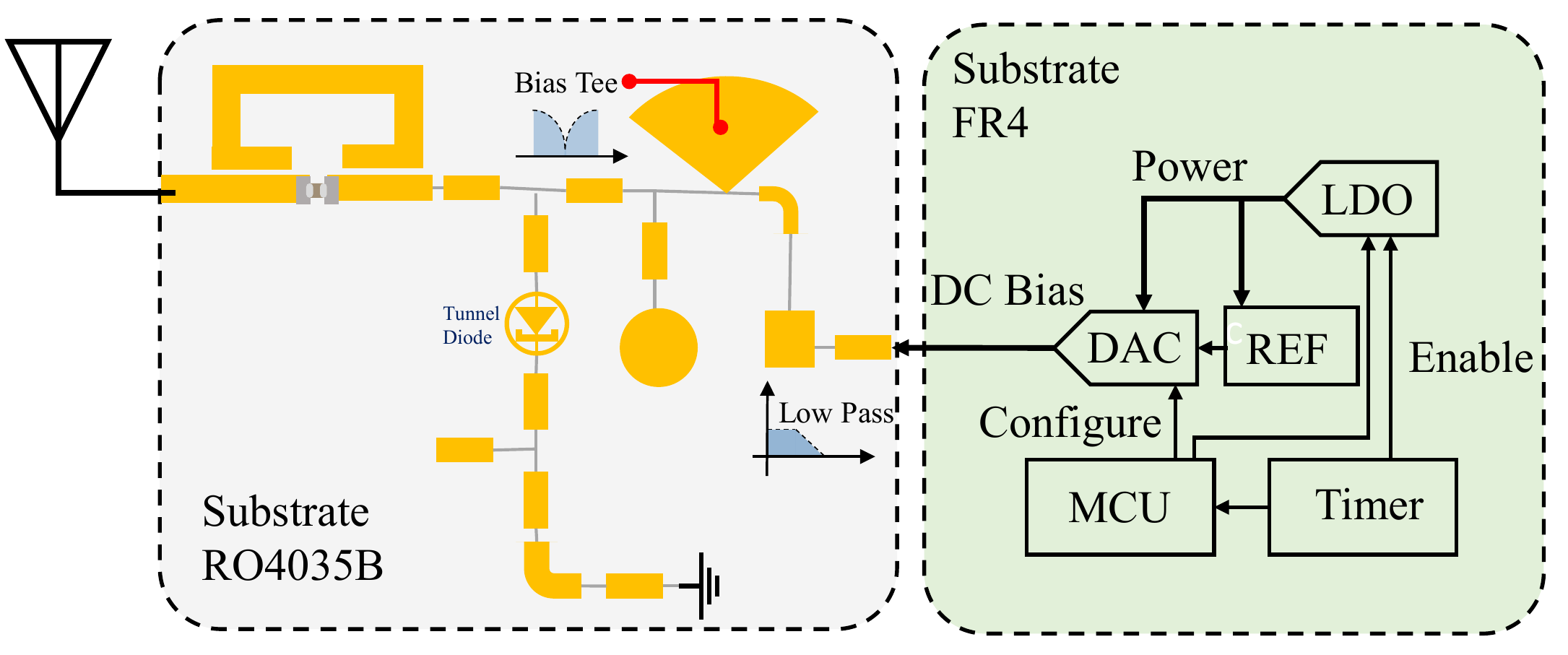}
		\end{minipage}%
	}%
	\subfigure[Verification]{
		\begin{minipage}[t]{0.35\linewidth}
			\centering
			\includegraphics[width=\linewidth]{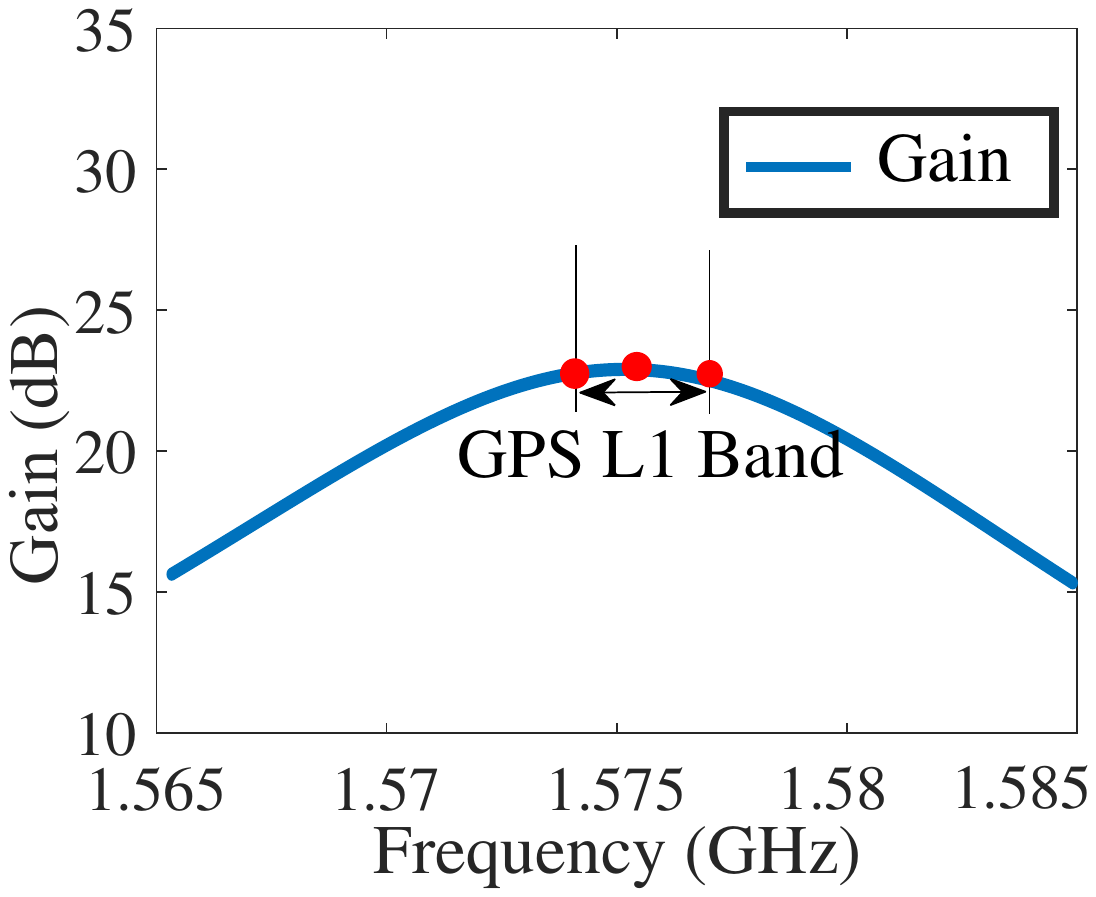}
		\end{minipage}%
	}%
	\caption{Schematic diagram and simulation result that verified the hardware performance.}
	\label{fig:antennaComparison}\vspace{-0.7cm}
\end{figure}	
\textbf{Operating with minimal noise.}
Before introducing the noise suppression design, we first describe the noise figure (NF) of the \sysname circuit.
NF determines the degradation of SNR of the output signal compared to the incident signal, which should be minimized to maintain good signal quality. 
Specifically, the NF of a tunnel diode is given by \cite{TunnelAmplifier,yariv1961noiseTunnelDiode}:
	\begin{small}
		\begin{equation} \label{equation:NF}
			\setlength{\abovedisplayskip}{3pt}
			\setlength{\belowdisplayskip}{3pt}
			NF=\frac{1+K_a}{\left[ 1-r/R_{NR}\right]\left[1- f/f_{r0} \right]  }
		\end{equation}
	\end{small}
where $r$ is the resistance of the Ge board as plotted in \fig~\ref{fig:model}, $f_{r0}$ the cutoff frequency, $f$ the operating frequency, $K_a\approx1.2$ the noise factor~\cite{TunnelAmplifier}, and $R_{NR}$ the negative resistance of the tunnel diode.

As shown in Figure~\ref{fig:IV-curve}, we can fit the relationship between the bias voltage $V_\text{Bias}$ and the current $I$ with a polynomial $I=\mathcal{F}(V_\text{Bias})$. 
Thus, there is a one-to-one mapping between the voltage and $R_{NR}$. We thus calculate all the NF values according to \eq~\ref{equation:NF} to obtain the minimal value of NF, as plotted in the \fig~\ref{fig:model}.

\section{GPS-Compatible Positioning through Backscatter Signals} 
Conventional GPS algorithms on smartphones may perform poorly in the shadowed regions due to the following two reasons. 
First, inadequate visible satellites, which are caused by the obstruction of buildings, make the traditional algorithm unsolvable. Conventional GPS algorithms, however, require receiving signals from at least four visible satellites in order to estimate four unknown variables $[x,y,z,t_b]$ simultaneously~\cite{googleGNSS}. Second, errors in distance measurements caused by NLoS propagation lead to inaccurate results.

Therefore, we develop a new algorithm that leverages the capabilities of \sysname tags to offer users the flexibility of prioritizing either extending locatable regions or improving accuracy.
In this section, we first clarify the characteristics of scattered GPS signals and then proceed to describe the details of our positioning algorithm.

\vspace{-0.3cm}
\subsection{Features of Scattered GPS Signals}\label{subsec:rawMeasurement}
Different from traditional GPS relay systems that just amplify GPS signals to enhance the coverage, we modulate sequences on GPS signals with the ``ON-OFF'' switching on the \sysname tag. These sequences can help distinguish the scattered and non-scattered signals.
Before elaborating on our algorithm design, we first introduce the accessible raw GPS measurements \sysname used~\cite{googleGNSS}.
\begin{figure}
	\centering
	\setlength{\abovecaptionskip}{0.cm}
	\includegraphics[width=0.8\linewidth]{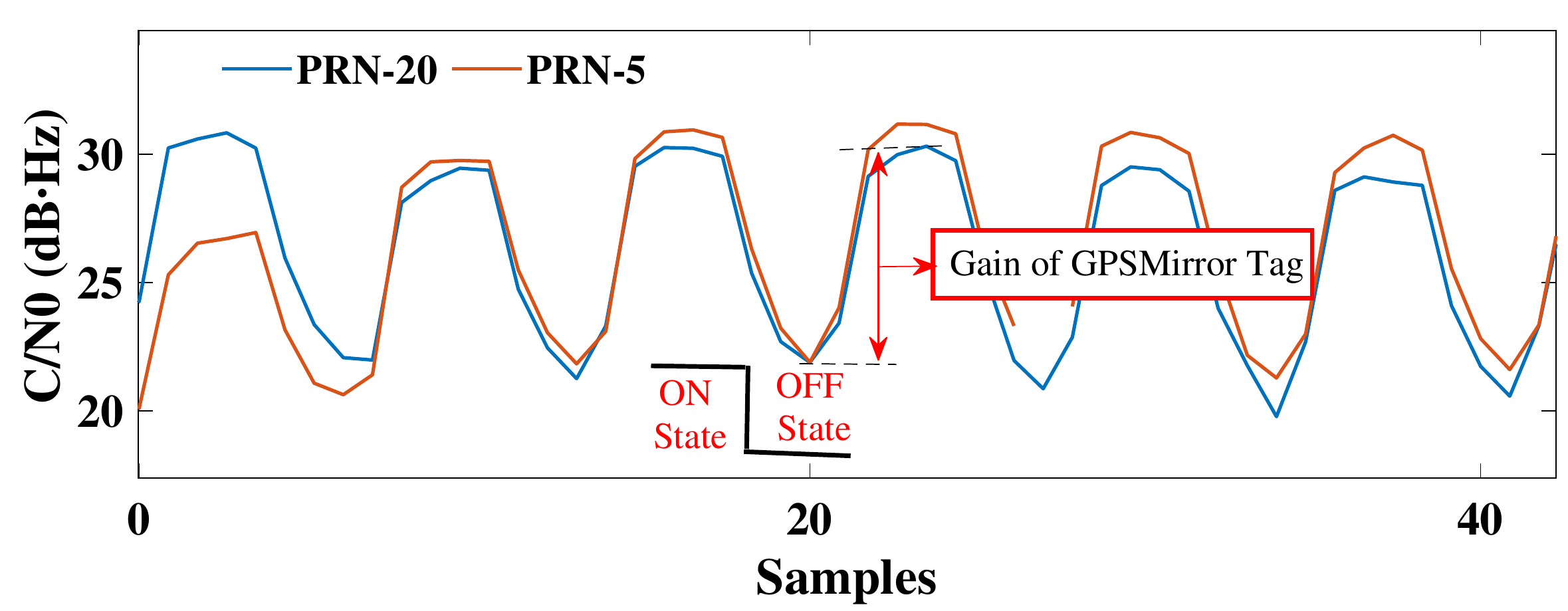}
	\caption{The amplitude feature of GPS signals scattered by a GPSMirror tag.}
	\label{fig:CN0Changes} \vspace{-0.7cm}
\end{figure}

\textbf{Signal strength measurements.}  
Smartphones provide the carrier-to-noise-density ratio ($C/N_{0}$) as the signal strength measurements to indicate the strength of the received GPS signal. 
Since the \sysname tags can enhance the GPS signals, the scattered GPS signals have higher $C/N_{0}$ measurements than the non-scattered signals of the same satellites. Therefore, we can modulate significant features on $C/N_{0}$ to determine which piece of signals of which satellites have been scattered.

To discriminate the tag and extract the signal that is scattered by a \sysname tag, we configure the tag periodically switching between ``ON'' and ``OFF'' states to generate a certain pattern for discrimination. During the ``ON'' state, the DAC provides bias voltage to the tunnel diode amplifier and the \sysname tag starts to re-radiate GPS signals with gain; whereas the ``OFF'' state shuts down the bias voltage so the \sysname tag keeps silent.
Such ON-OFF keying can significantly change the measured signal strength of GPS signals and thus changes the $C/N_{0}$ measurements on smartphones. 

For a better understanding, we use a smartphone to record the $\mathbf{C/N_{0}}$ samples with a \sysname tag deployed 1~m away. Plotted in \fig~\ref{fig:CN0Changes} is the measured $\mathbf{C/N_{0}}$ from GPS satellite PRN-20 and PRN-5. The $\mathbf{C/N_{0}}$ changes significantly with the \sysname tag switching from the ``ON'' state to the ``OFF'' state. We define the ratio of the $\mathbf{C/N_{0}}$ between the ``ON'' state and ``OFF'' state is the \textit{gain} of \sysname, and use this value to evaluate the coverage of \sysname tags in \S~\ref{sec:evalCoverage}.

\begin{figure}
	\centering
	\setlength{\abovecaptionskip}{0.cm}
	\subfigure{
		\begin{minipage}[t]{0.4\linewidth}
			\centering
			\includegraphics[width=\linewidth]{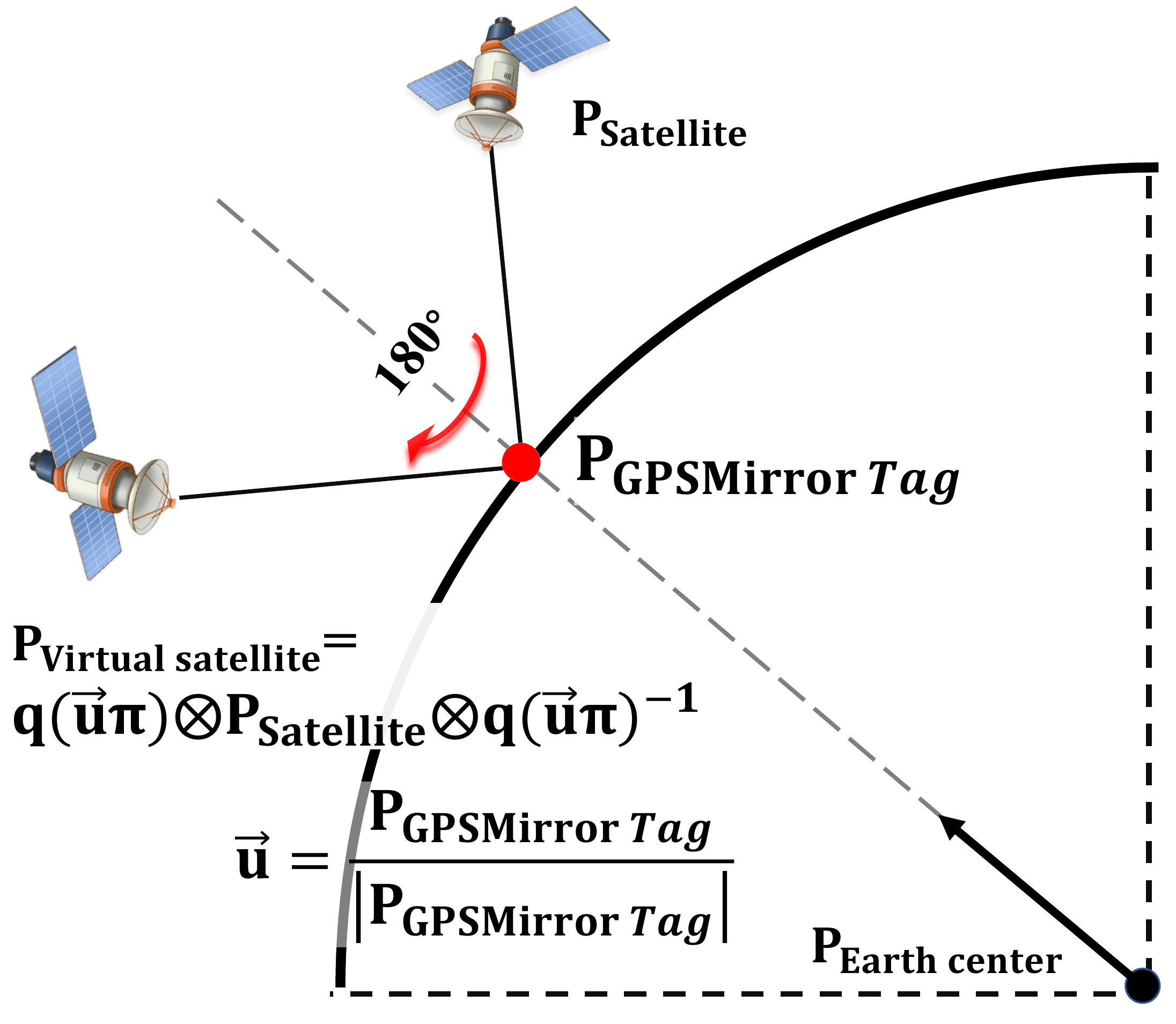}
		\end{minipage}%
	}%
	\subfigure{
		\begin{minipage}[t]{0.4\linewidth}
			\centering
			\includegraphics[width=\linewidth]{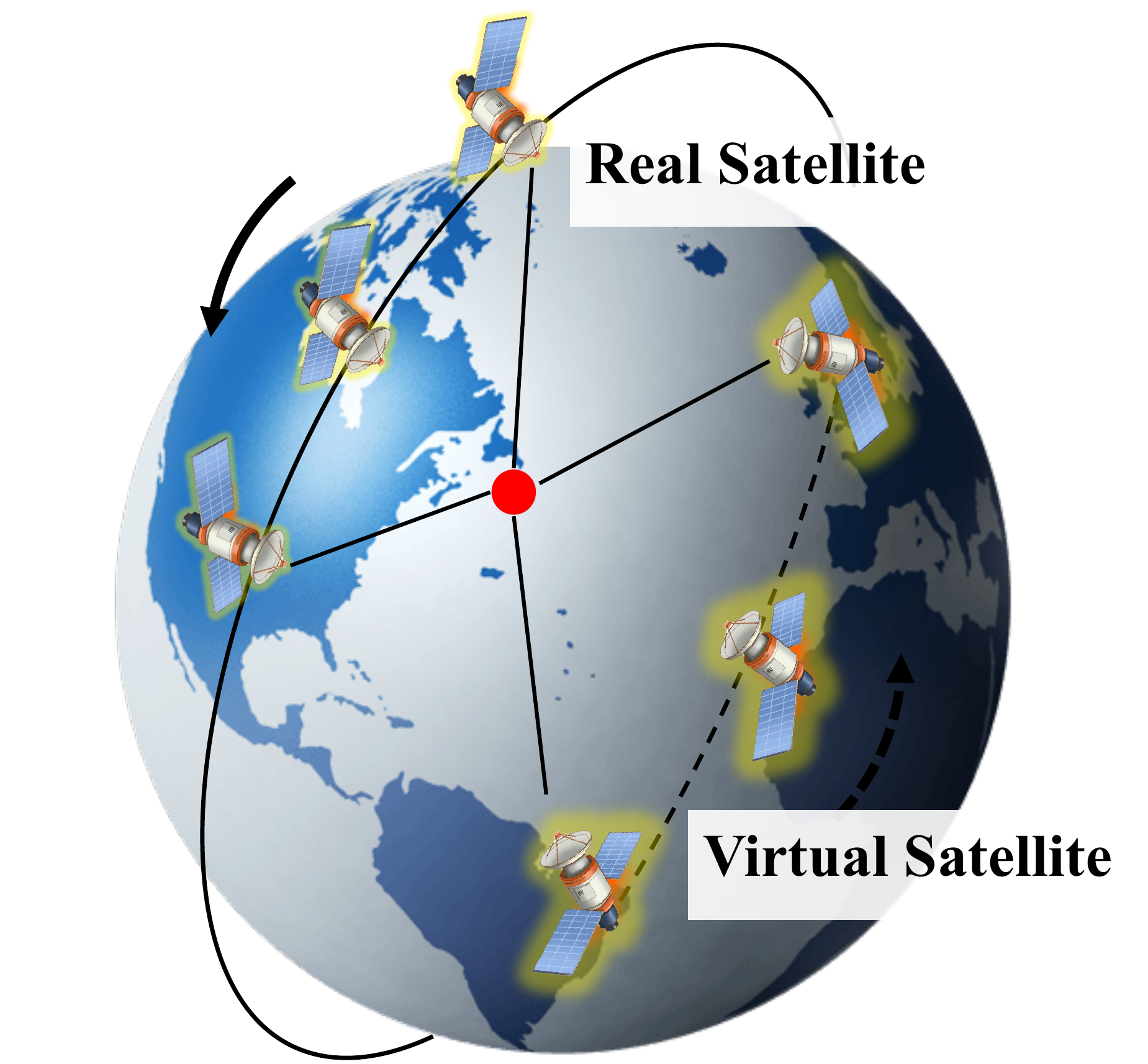}
		\end{minipage}%
	}%
	\caption{Position estimation of virtual satellites.}
	\label{fig:VirtualSat}\vspace{-0.7cm}
\end{figure}

\textbf{Distance measurements.}
Smartphones provide two types of measurements related to the distance between the satellites and the smartphone: the \textit{Pseudorange} and the \textit{Accumulated Delta Range} measurements. The \textit{Pseudorange} is a time-of-flight measurement calculated using the local clock of the smartphone, while the \textit{Accumulated Delta Range} is a count of carrier cycles since the GPS sensor was started. Both measurements contain bias errors and cannot be directly used for positioning. Since we deploy \sysname to re-radiate GPS signals with ``ON-OFF'' scattering, the scattered GPS signals have additional delay $t_s$ which is not contained in conventional GPS algorithm~\cite{IoNhumphreys2016feasibility}.  

Pseudorange $\rho$ of scattered GPS signals can be modeled as
\begin{equation}
	\setlength{\abovedisplayskip}{3pt}
	\setlength{\belowdisplayskip}{3pt}
	\rho=r+ct_b+ct_s+\varepsilon_{\rho} \label{eqution:pseudorange}
\end{equation}
where $t_b$ is the clock bias between a satellite and a receiver, which is the main bias error in GPS distance measurements and needs to be correctly estimated to extract accurate distance. $r$ is the distance between the smartphone and the GPS satellite without error, $c$ is the speed of light, and $\varepsilon_{\rho}$ is the measurement error caused by other uncertain factors including ionospheric delay $I$, tropospheric delay $T$, etc. The scattered GPS signals have an additional delay $t_s$ while $t_s=0$ for the non-scattered signals.  
The distance measurements from accumulated carrier phase $\phi$, denoted as $\Phi=\lambda\phi$, can typically be extracted from \textit{Accumulated Delta Range} measurements. The relationship between $\Phi$ and $r$ can be expressed by
\begin{small}
	\begin{equation}\label{Eq:Phi}
		\setlength{\abovedisplayskip}{3pt}
		\setlength{\belowdisplayskip}{3pt}
		\Phi= r + ct_b +ct_s+\lambda N +\varepsilon_{\Phi} 
	\end{equation}  
\end{small}
where $\lambda$ is the wavelength of the GPS signals, $N$ the phase offset relative to the initial reference phase value (commonly called ambiguity of whole cycles), and $\varepsilon_{\Phi}$ the measurement errors.
Note that $N$ is fixed since the carrier phase is kept locked by the receiver and can be estimated using a double difference of pseudorange measurements between satellites. We can estimate  $N$ by $N\approx[(\Phi-\rho)/\lambda]$~\cite{kayton2002global}.

\vspace{-0.4cm}
\subsection{Extending Locatable Regions under Inadequate Satellites}
Under inadequate satellite conditions ($<$4), the measurable propagation paths for localization are insufficient to estimate the position and clock bias. Hence, to turn these regions into locatable regions, we use \sysname tags to create additional measurable propagation paths through ``ON-OFF'' switching. To make the position estimation algorithm solvable, we then extract propagation delays through a \sysname tag and create ``virtual satellites'' that matched these measurable propagation paths. Finally, we estimate the position with combined measurements.

\textbf{Estimating the position of virtual satellites.} 
To correctly solve \eq~(\ref{eqution:pseudorange}), the virtual satellites must comply with the following rules. (i) A virtual satellite and the real satellite that emits the scattered GPS signals need to share the same pseudorange and Doppler shift to ensure the correctness of the solution. (ii) The orbit of a virtual satellite is still a true GPS orbit. (iii) A virtual satellite needs to be far away from the real satellite to ensure a small Dilution of precision (DOP) and make the positioning results robust to measurement uncertainty.
To meet such requirements, we create virtual satellites by rotating the original satellites' position vectors by 180$^{\circ}$ centered on the \sysname node, as shown in Figure~\ref{fig:VirtualSat}. In particular, we create a virtual satellite with coordinates $P_\text{vs}$ from a real satellite with coordinates $P_\text{rs}$ according to the standard rotation equation~\cite{altman1986rotations}.

\textbf{Extracting the delay of scattered signals.}
The propagation delay, $t_s$, through a \sysname tag in \eq~\eqref{eqution:pseudorange} must be determined prior to positioning. Given that both scattered and non-scattered GPS signals are received by the same smartphone, the clock bias remains constant. Assuming we obtain nearby static samples of GPS signals and extract clock bias $t_{b1}$ from non-scattered signals, we can assume the clock bias from scattered signals $t_{b2}$ is equal to $t_s+t_{b1}$. By continuously comparing scattered and non-scattered GPS measurement pairs, we can estimate and update $t_s$.

\textbf{Positioning the receiver with a single \sysname tag.}
To estimate the receiver's position, we use the weighted least squares (WLS) to combine the measured pseudorange directly from scattered and non-scattered GPS signals. The process of estimating the receiver's position through WLS involves setting an initial position $L_0=[x_0,y_0,z_0]=[0,0,0]$ to activate the iterative algorithm. We then obtain the locations of the tracked satellites from the ephemeris, which provides GPS satellite's orbital information forecasted by NASA.

Different from conventional algorithms, \sysname needs to calculate the distance from both the real and virtual satellites to the given initial position. Assuming $m$ satellite signals, including real and virtual satellites, are to be acquired. For the $k$-th satellite ($k\leq m$), using $L^{(k)}=[x_k,y_k,z_k]$ to represent its location and $t_{b0}$ as the initial clock bias, then the true distance $r$ in the Equation~(\ref{eqution:pseudorange}) can be derived as $||L^{(k)}-L_{0}||$, and the measured pseudorange $\rho_0^{(k)}$ follows 
\begin{small}
	\begin{equation}
		\setlength{\abovedisplayskip}{3pt}
		\setlength{\belowdisplayskip}{3pt}
		\rho_0^{(k)}=|| L^{(k)}- L_{0} ||+ ct_{b0} +\varepsilon_{\rho}
	\end{equation}
\end{small}
Next, we linearize the above equation. Assuming that the true position is $L = L_0 + \delta L$, the difference between the target position (true position) and the initial position can be derived as
\begin{small}
	\begin{equation} \label{eq:linealy}
		\setlength{\abovedisplayskip}{3pt}
		\setlength{\belowdisplayskip}{3pt}
		\begin{aligned}
			\delta \rho^{(k)} &= \rho^{(k)}-\rho_{0}^{(k)}  \\
			&= || L^{(k)}- (L_0 + \delta L) || - || L^{(k)}- L_0 || + c(t_{b}-t_{b0})+ \varepsilon_{\rho}\\
			&\approx -\frac{ L^{(k)} - L_0}{ L^{(k)}- L_0} \delta L +c\delta t_{b} + \varepsilon_{\rho}=-I^{(k)} \delta L + c\delta t_{b} + \varepsilon_{\rho}
		\end{aligned}
	\end{equation}
\end{small}
which shows that $\delta \rho^{(k)}$ and $\delta L$ are linearly related. Therefore, both pseudorange measurements from real satellites and virtual satellites can be linearized using \eq~\eqref{eq:linealy}. 

Different from conventional GPS algorithms which simply apply the WLS solver with these measurements, the position estimation equations are reconstructed with the extracted delay of scattered signals. 
Specifically, suppose we obtain two consecutive pseudorange measurements, $\rho^{(k,n)}$, from non-scattered GPS signals, and $\rho^{(k,s)}$, from \sysname, which can be regarded as the signal from the virtual satellites.  Let $\delta t_{b1}$ be the clock bias derived directly from the satellites and $\delta t_{b2}=\delta t_{b1}+ t_{s}$ be the estimated clock bias, where $\delta t_{b2}$ is the true clock bias and $t_{s}$ is the propagation delay through the \sysname node. If we obtain $k$ measurements $\textbf{P}_n=[ \rho^{(1,n)},\cdots,\rho^{(k,n)}]^T -\rho_0$ directly from the satellite and $m$ measurements $\textbf{P}_s=[\rho^{(1,s)},\cdots,\rho^{(m,s)}]^T -\rho_0$ from the \sysname node, We can organize the pseudorange of all measurements as
\begin{small}
	\begin{equation}
		\setlength{\abovedisplayskip}{3pt}
		\setlength{\belowdisplayskip}{3pt}
		\delta \rho= \left[ \begin{array}{ccc}
			\textbf{P}_n\\
			\textbf{P}_s
		\end{array} 
		\right ] = \left[ \begin{array}{ccc}
			-\textbf{I}_n^T&1&0 \\
			-\textbf{I}_s^T&0&1 \\
		\end{array} \right]\left[\begin{array}{ccc}
			\delta L \\
			c\delta t_{b1} \\
			c\delta t_{b2}
		\end{array}\right] + \varepsilon_{\rho},
	\end{equation}
\end{small}
Then, we can estimate the $\delta L$ through WLS:
\begin{small}
	\begin{equation} \label{equation:WLS}
 		\setlength{\abovedisplayskip}{3pt}
		\setlength{\belowdisplayskip}{6pt}
		\left[ \begin{array}{ccc}
			\delta  L \\
			c\delta  t_{b1} \\
			c\delta  t_{b2}
		\end{array}\right] = ( (G^TWG)^{-1}G^TW \delta  \rho)
	\end{equation}
\end{small}
where $W$ is the weight of each equation from the measurement uncertainties and $G$ denote the Jacobian matrix calculated using satellite positions.
We can determine the true position as $L = L_0 + \delta L$. 
As $t_s$ can be extracted by comparing scattered and non-scattered GPS signal pairs originating from the same satellites, a single \sysname tag may enable smartphone positioning with merely two visible satellites after obtaining $t_s$. Additionally, accuracy can be improved with the following approaches if the number of visible satellites exceeds 2.

\vspace{-0.3cm}
\subsection{Differential Positioning for Accuracy Improvement}

\begin{figure}[t] 
	\centering
	\setlength{\abovecaptionskip}{0.cm}
 	\setlength{\belowcaptionskip}{0.cm}
		\begin{minipage}[t]{0.4\linewidth}
			\centering
			\includegraphics[width=\linewidth]{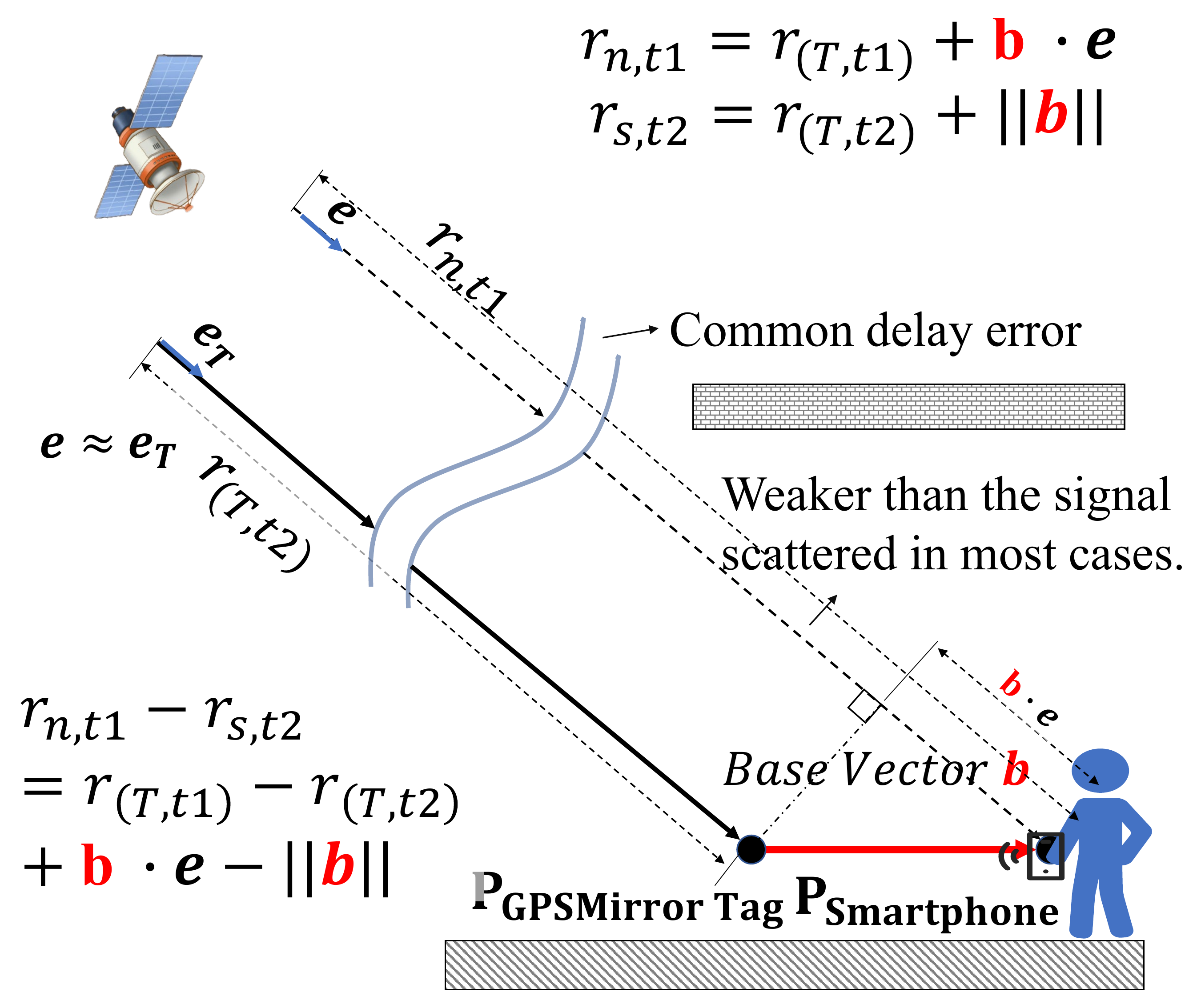}
		\end{minipage}%
		\begin{minipage}[t]{0.55\linewidth}
			\centering
			\includegraphics[width=\linewidth]{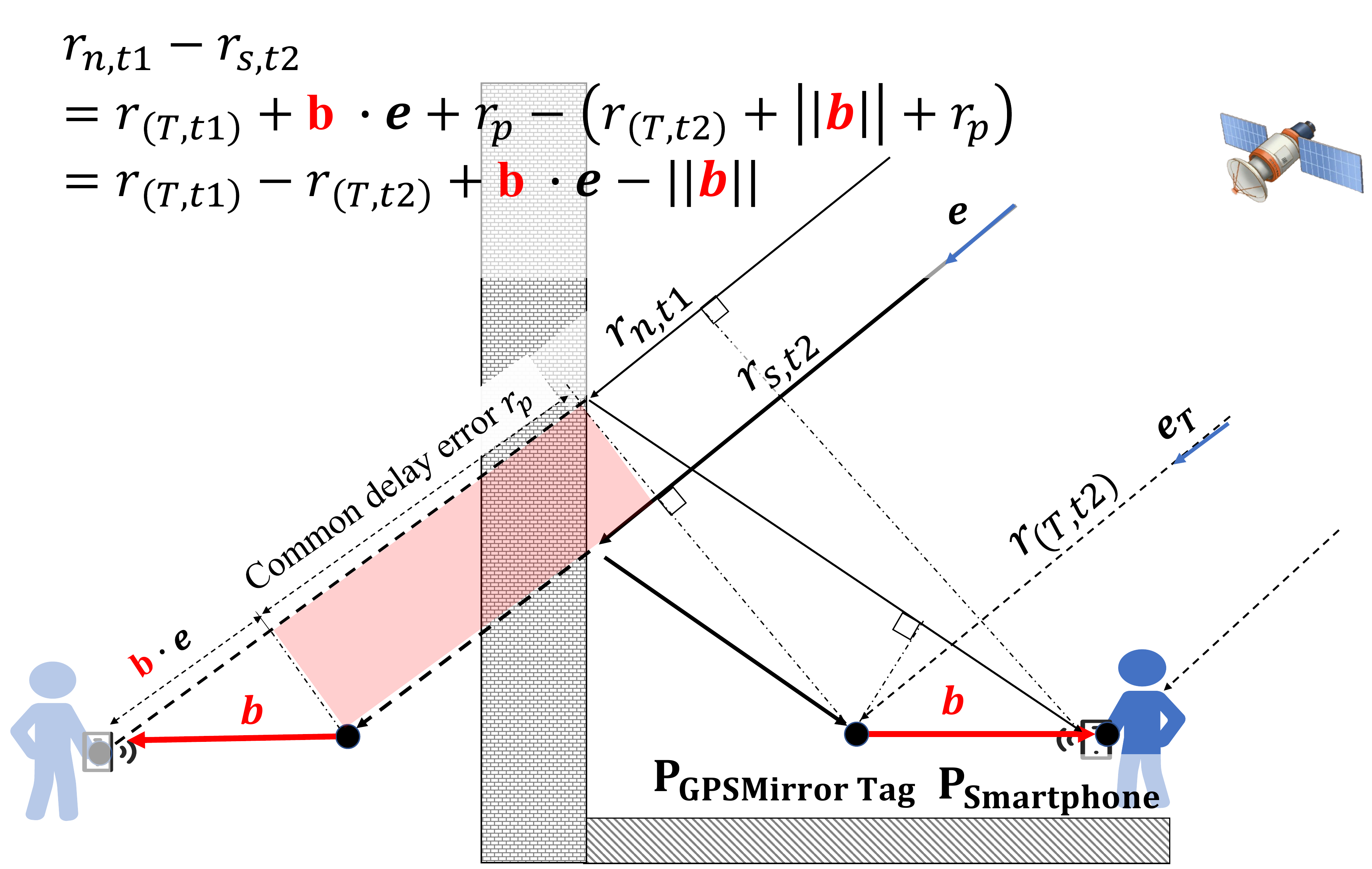}
		\end{minipage}%
	\caption{Typical use cases for a \sysname tag to enhance the GPS positioning in shadowed regions.}
	\label{fig:Raw_Phi_measurement}
 \vspace{-0.7cm}
\end{figure}
The other problem in shadowed regions is NLoS propagation. Intuitively, the lack of LoS causes rapid degradation in positioning accuracy. Since the LoS from the satellites to \sysname tags and LoS from the \sysname tags to smartphones are relatively easy to ensure, an important question is: can we convert the legacy positioning method that requires LoS from the receiver to satellites into a positioning method that requires LoS to a \sysname tag?

Our basic idea is to cast the \sysname tags as the reference stations, which create deterministic multipaths to improve positioning accuracy and mitigate the errors caused by random ambient reflections.
Unlike conventional reference station-based systems (e.g., DGPS~\cite{morgan1995differential}) where the reference stations can process and compute the propagation delays correctly and deliver them to smartphones, the low-power \sysname tags with limited resources cannot process GPS signals or determine the propagation delays. Thus, \sysname needs to extract distance measurements and complete the computation using smartphones. 
Since the satellites are far from the ground (about 20000~km), scattered and non-scattered GPS signals from the same satellite in a short interval can be considered as arriving in parallel with the common propagation errors, which are mainly caused by NLOS in shadowed regions. We cast the \sysname tags as reference stations to reduce these errors.

Distinct from DGPS stations, \sysname tags are incapable of directly processing GPS signals and extracting errors from the referenced propagation path. It is necessary to process signals from the referenced propagation path and eliminate propagation errors on smartphones.
As depicted in \fig~\ref{fig:Raw_Phi_measurement}, a \sysname tag is employed to scatter GPS signals to a smartphone and differentiate the distance measurements of scattered and non-scattered GPS signals, thereby eliminating the common delay error. Pseudorange and phase measurements, both applicable for differential purposes, display comparable mathematical expressions. Owing to the superior resolution, phase measurements take precedence for positioning. But in some cases where phase data is unstable, pseudorange measurements can also be employed. The following delineates the detailed steps. Assuming two proximate samples of carrier phase measurements are obtained, $\Phi_{n,t1}$ from non-scattered GPS signal at $t1$ and $\Phi_{s,t2}$ from scattered signals of the same satellite at $t2$. During this short interval, the angle of arrived GPS signal changes negligibly~\cite{parkinson1996differential}. Consequently, it can be assumed that proximate samples of scattered and non-scattered GPS signals share identical common errors, encompassing the delay induced by reflection and atmospheric delays $I$ and $T$. Furthermore, differentiating measurements from scattered and non-scattered GPS signals aids in eliminating common errors that undermine positioning accuracy. Explicitly, $\Phi_{d}=\Phi_{s,t2}-\Phi_{n,t1}$ can be acquired, where $\Phi_{s,t2}$ is the distance measurements from scattered GPS signals and $\Phi_{n,t1}$ from non-scattered GPS signals of the same satellites. According to \eqref{Eq:Phi}, the equation can be expressed by
\begin{small}
	\begin{equation}
 	\setlength{\abovedisplayskip}{3pt}
	\setlength{\belowdisplayskip}{3pt}
		\Phi_{d}=r_{s,t2}-r_{n,t1}+c\Delta T +\Delta N + \varepsilon_{\Phi} 
	\end{equation}
\end{small}
where $\Delta N$ is the difference of the phase offset and remains constant when the carrier phase is locked, both $r_{s,t2}$ and $r_{n,t1}$ are the distance between the same satellite to the smartphone at a different time. $\Delta T = (t_{R,t2}-t_{R,t1})- (t_{S,t2}-t_{S,t1})$ is the difference of clock bias between the 2 measurements. As the clocks of satellites are stable, $\Delta T$ mainly depends on the stability of the local clock of smartphones. 

Different from conventional DGPS stations, \sysname cannot estimate the $\Phi_{d}$. Thus, we model an unknown vector $\textbf{b}$ to stand for a position vector from the tag to the smartphone and estimate it on the smartphone. As shown in \fig~\ref{fig:Raw_Phi_measurement}, we use $r_{T,t2}$ to denote the distance from the satellite to the \sysname tag from scattered GPS signals. Then we can presume $r_{s,t2}=r_{T,t2}+||\textbf{b}||$ based on the geometric relationship in \fig~\ref{fig:Raw_Phi_measurement}. Further, we can acquire $r_{n,t1}=r_{T,t1}+\textbf{b}\cdot \textbf{e}$, where $e$ is the unit directional vector from the satellite to the smartphone and $\Phi_{d}$ can be expressed as 
\begin{small}
	\begin{equation} 
 	\setlength{\abovedisplayskip}{3pt}
	\setlength{\belowdisplayskip}{3pt}
		\begin{aligned} 
			\Phi_{d}=\Delta r_T+||\textbf{b}||-\textbf{b}\cdot \textbf{e} +c\Delta T +\lambda \Delta N + \varepsilon_{\Phi} \\
		\end{aligned} 
	\end{equation} 
\end{small}
where $\Delta r_T =r_{T,t2}-r_{T,t1}$ is the change in distance from the satellites to the \sysname between $t1$ and $t2$. We have established the relationship between the measurements and the base vector $\textbf{b}$ and reorganized the equation by placing the known quantities on the left side.
\begin{small}
	\begin{equation} \label{Eq:Phi_DiffObservable}
  	\setlength{\abovedisplayskip}{3pt}
	\setlength{\belowdisplayskip}{3pt}
		\begin{aligned}
			\Phi_{d}-\Delta r_T-\lambda \Delta N &= ||\textbf{b}||-\textbf{b}\cdot \textbf{e} +c\Delta T + \varepsilon_{\Phi} \\ 
			&= \sqrt{\textbf{b}^T\textbf{b}}-\textbf{b}\cdot \textbf{e} +c\Delta T + \varepsilon_{\Phi}
		\end{aligned}
	\end{equation}  
\end{small}
Given a $\textbf{b}_0$ for initialization, the Equation~\eqref{Eq:Phi_DiffObservable} can be linearized by
\begin{small}
	\begin{equation}  \label{Eq:Phi_linear}
   	\setlength{\abovedisplayskip}{3pt}
	\setlength{\belowdisplayskip}{3pt}
		\begin{aligned}
			\Phi_{d}-\Delta r_T-\lambda \Delta N &\approx \sqrt{\textbf{b}_0^T\textbf{b}_0}-\textbf{b}_0\cdot \textbf{e} +\frac{\partial (\sqrt{\textbf{b}^T\textbf{b}}-\textbf{b}\cdot \textbf{e})}{\partial \textbf{b}}\delta \textbf{b} \\
			&+c\Delta T + \varepsilon_{\Phi}  \\ 
		\end{aligned}
	\end{equation}
\end{small}
with Newton's iterative method. We use $S_\Phi=\Phi_{d}-\Delta r_T-\lambda \Delta N - \sqrt{\textbf{b}_0^T\textbf{b}_0}+\textbf{b}_0\cdot \textbf{e}$ to represent the known quantities. Equation~\eqref{Eq:Phi_linear} can be rearranged to
\begin{small}
	\begin{equation} \label{Eq:States_Linear}
        \setlength{\abovedisplayskip}{3pt}
	\setlength{\belowdisplayskip}{3pt}
	S_\Phi=\frac{\partial (\sqrt{\textbf{b}^T\textbf{b}}-\textbf{b}\cdot \textbf{e})}{\partial \textbf{b}}\delta \textbf{b} +c \Delta T + \varepsilon_{\Phi}
	\end{equation}
\end{small}
Now according to \eq~\eqref{Eq:States_Linear}, the measured phase of GPS signals can be simply determined by the $b$ from the \sysname tag to the smartphone, which eliminates the requirement for LoS satellites. An intuitive explanation is that
since the tag's position is known and the position of satellites can also be calculated based on the orbit information, the propagation path between satellites and the \sysname tag is derived indirectly.

Then we start to solve the base vector $b$ with various formulas. We can obtain the base vector $\textbf{b}$ by estimating the $\delta \textbf{b}$ since they have a linear relationship. There are 4 variables in Equation~\eqref{Eq:States_Linear}, $\textbf{b}=[\textbf{X}_{Smartphone}-\textbf{X}_{Tag},\textbf{Y}_{Smartphone}-\textbf{Y}_{Tag},\textbf{Z}_{Smartphone}-\textbf{Z}_{Tag}]$ and $\Delta T$. Thus, the system has sufficient equations to solve $\textbf{b}$. 
After we obtain $n \geq 4$ equations, we have 
\begin{small}
	\begin{equation}
   	\setlength{\abovedisplayskip}{3pt}
	\setlength{\belowdisplayskip}{3pt}
		\textbf{S}_\Phi= \left[ \begin{array}{c}
			S_\Phi^{(1)}\\
			\vdots\\
			S_\Phi^{(n)}\\
		\end{array} 
		\right ] = 
		\left[\begin{array}{cc}
			\frac{\partial \sqrt{\textbf{b}^T\textbf{b}}-\textbf{b}\cdot \textbf{e}}{\partial \textbf{b}} & 1 \\
			\vdots & \vdots \\
			\frac{\partial \sqrt{\textbf{b}^T\textbf{b}}-\textbf{b}\cdot \textbf{e}}{\partial \textbf{b}} & 1 
		\end{array}\right] 	
		\left[ \begin{array}{c}
			\delta \textbf{b} \\
			c\Delta T \\
		\end{array} \right]+ \varepsilon_{\Phi},
	\end{equation}
\end{small}
Then, we can estimate the $\delta b$ using a WLS estimation equation as
\begin{small}
	\begin{equation} \label{equation_WLS}
 		\setlength{\abovedisplayskip}{3pt}
		\setlength{\belowdisplayskip}{3pt}
		\left[ \begin{array}{c}
			\delta \textbf{b} \\
			c\Delta  T
		\end{array}\right] = ( (G^TWG)^{-1}G^TW \textbf{S}_\Phi)
	\end{equation}
\end{small}
We then update $\textbf{b}_0=\textbf{b}_0 +\delta \textbf{b}$ and iterate until the $\delta \textbf{b}$ is lower than a preset threshold. Then we can obtain the estimated base vector $\textbf{b}$. Further, since the location of \sysname tags and the floor plan is known after deployment, we filter out outlier $\textbf{b}$ with impossible directions to further improve accuracy. 
Finally, we obtain the position of the smartphone with $P_{Smartphone}=P_{GPSMirror Tag} + \textbf{b}$, demonstrated in \fig~\ref{fig:Raw_Phi_measurement}. Moreover, we can refine the direction and position of $\textbf{b}$ based on the floor plan of the deployment site, thereby filtering out obvious incorrect directions to achieve higher accuracy.

%
\section{System Implementation} \label{sec: HardwareImplementation} 
In this section, we describe the implementation of \sysnamenospace, including fabrication of \sysname tags, and GPS data logger on receivers, i.e.,smartphones.
 \vspace{-0.3cm}
\begin{figure}[t] 
	\centering
	\setlength{\abovecaptionskip}{0.cm}
	\subfigure[RF front-end of \sysname tags.]{
		\begin{minipage}[t]{0.67\linewidth}
			\centering
			\includegraphics[width=\linewidth]{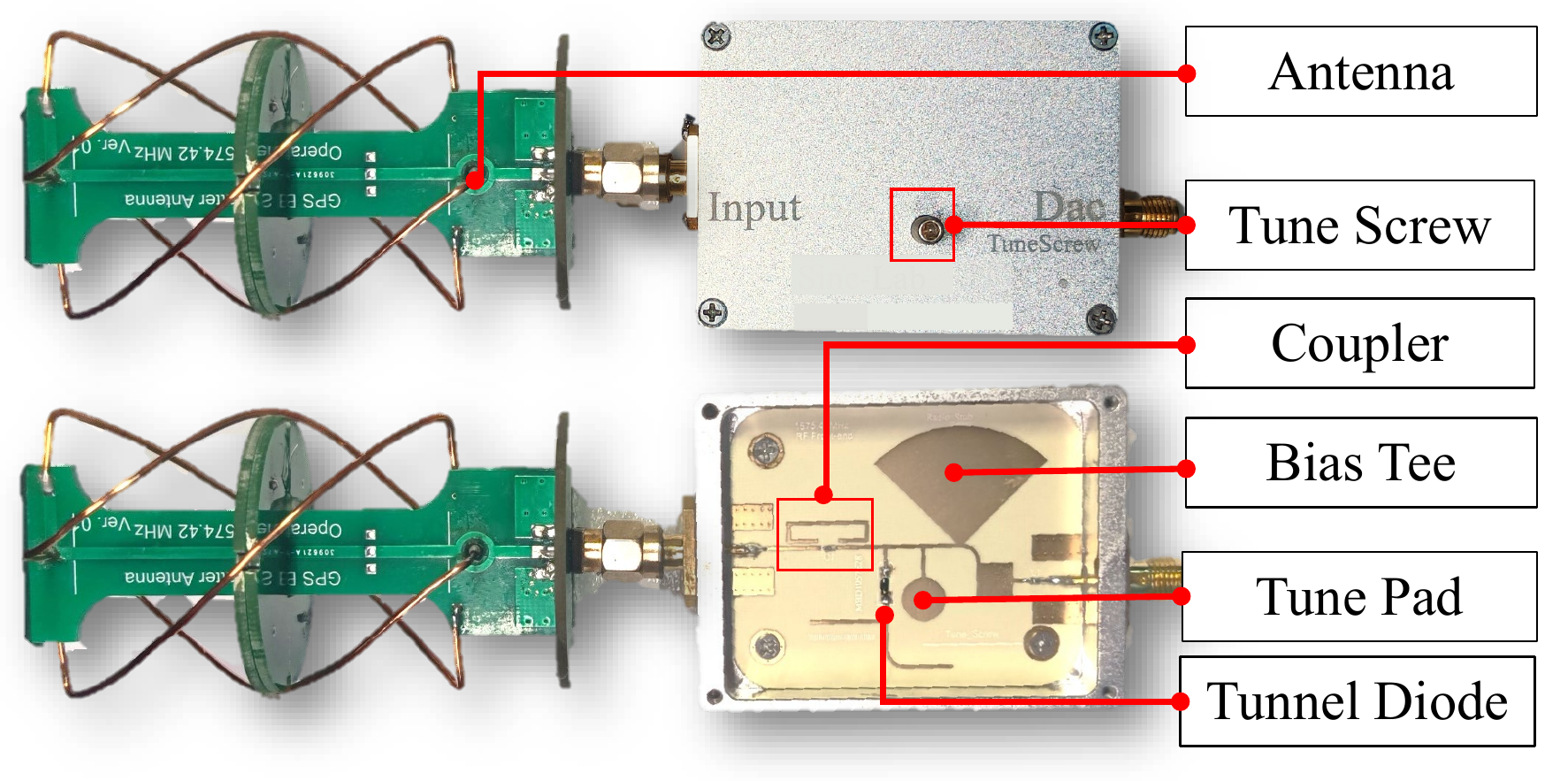}
		\end{minipage}%
	}%
	\subfigure[Control circuit that provide bias voltage.]{
		\begin{minipage}[t]{0.3\linewidth}
			\centering
			\includegraphics[width=\linewidth]{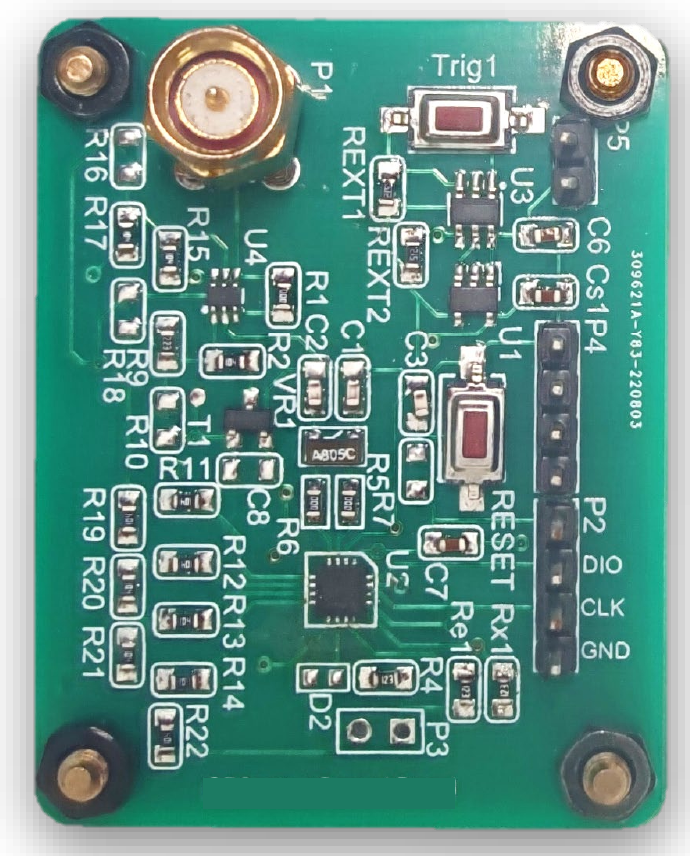}
		\end{minipage}%
	}%
	\caption{\sysname prototype.}
	\label{fig:prototype}\vspace{-0.5cm}
\end{figure}

\subsection{\sysname Tags} 
To provide accurate, stable and configurable bias voltage for the tunnel diode with off-the-shelf components, we prototype each \sysname tag with the following components: a tunnel diode to provide negative resistance for GPS signal amplification, a DAC to provide accurate bias voltage for the tunnel diode, a voltage reference chip that provides voltage reference against voltage-drift of the circuit, an MCU that configures and controls other components, a voltage buffer that holds the bias when other components are shut down to save energy and a low dropout regulator (LDO) that provides energy for the above components. 

To facilitate manufacturing and commissioning, we split the above components into two circuits. One is the front-end circuit that handles the RF signals with a tunnel diode and micro-strip lines, the other is the baseband control circuit that configures the bias voltage for the tunnel diode.  
As shown in Figure~\ref{fig:prototype}~(a), we fabricate the front-end circuit on a low-loss Rogers 4350B substrate with an immersion gold craft to reduce the impact of surface roughness and oxidation on impedance. Each circuit is mounted in a customized metal cavity to avoid ambient electromagnetic noise. Shown in Figure~\ref{fig:prototype}~(b) is the baseband control circuit fabricated on FR4.
In addition, we also build the customized antenna with a FR4 printed circuit board and copper wires.
\begin{tiny}
	\begin{table}[h]
		\centering
		\setlength{\abovecaptionskip}{0.cm}
		\setlength{\belowcaptionskip}{0.cm}
		\caption{GPSMirror Cost Breakdown.}
			\begin{tabular}{| c | c | c | }
				\hline \rowcolor{gray!40}
				Components & Model &  Price \\ 
				\hline
				Tunnel Diode & MBD1057 &  \$45.2 \\  
				\hline
				MCU & KL03Z16 & \$2.99  \\  
				\hline
				Timer & TPL5111DDCR &  \$0.30  \\  
				\hline
				Voltage Buffer & LTC2063SC6 & \$1.38  \\  
				\hline
				Voltage Reference & REF3012 & \$0.45  \\  
				\hline
				DAC & MAX5531 & \$4.34  \\  
				\hline
				LDO & TPS78218DDCT & \$0.50 \\  
				\hline
				Total & -- &\$55.16  \\  
				\hline
			\end{tabular}
		\label{tab:componentCost} \vspace{-0.4cm}
	\end{table}
\end{tiny}

\textbf{Cost.} Table~\ref{tab:componentCost} is the latest detailed breakdown of cost based on quotes for 1000 units, where the unit cost is around $\$ 55.16$, but only $\$ 9.96$ if excluding the tunnel diode. Tunnel diodes contribute to 82\% of the total cost owing to the lack of market demand. If \sysname were successfully promoted and widely deployed, mass production would further lower the unit price of tunnel diodes.

\textbf{Power consumption.}
The tunnel diode consumes approximately 126 $\mu$W according to \fig~\ref{sec:hardware} and the off-the-shelf LDO consumes approximately 1.6 $\mu$W. To reduce overall energy consumption, the MCU and DAC operate duty-cycled to provide precise voltage output and the voltage buffer helps to hold the bias voltage for the tunnel diodes when the MCU and DAC are shut-down. The scattering control and configuration of the DAC can be finished within 10~ms, thus the average power consumption of all these components is as low as 0.9 $\mu$W. The voltage buffer consumes 6 $\mu$W and the total power consumption is approximately 134.5~$\mu$W. 
134.5~$\mu$W-power-consumption can enable the \sysname to continuously work for about 31 months with a coin-battery CR2477, which is also used in duty-cycled iBeacon~\cite{IBeacon} for a maximum of 24 months, while GPS relays/repeaters consume W-level power and require a plug-in supply. 
In addition, since distance measurements can be derived from any 1~ms-long GPS signal~\cite{kayton2002global}, the tunnel diode can also operate in duty-cycled mode to amplify GPS signals. \sysname tags can achieve battery-free operation with RF or PV energy harvesters like other backscatter systems.

 \vspace{-0.3cm}
\subsection{Receiver} 
We use various smartphones including Xiaomi 8, Huawei Mate20 Pro and Samsung Galaxy S20 Ultra with the GnssLogger 3.0~\cite{googleGNSS} app installed for recording data. We process the recorded data and calculate the location offline.
\begin{figure}[t]
	\vspace{0pt}
	\centering
	 \setlength{\abovecaptionskip}{0.cm}
	\includegraphics[width=\linewidth]{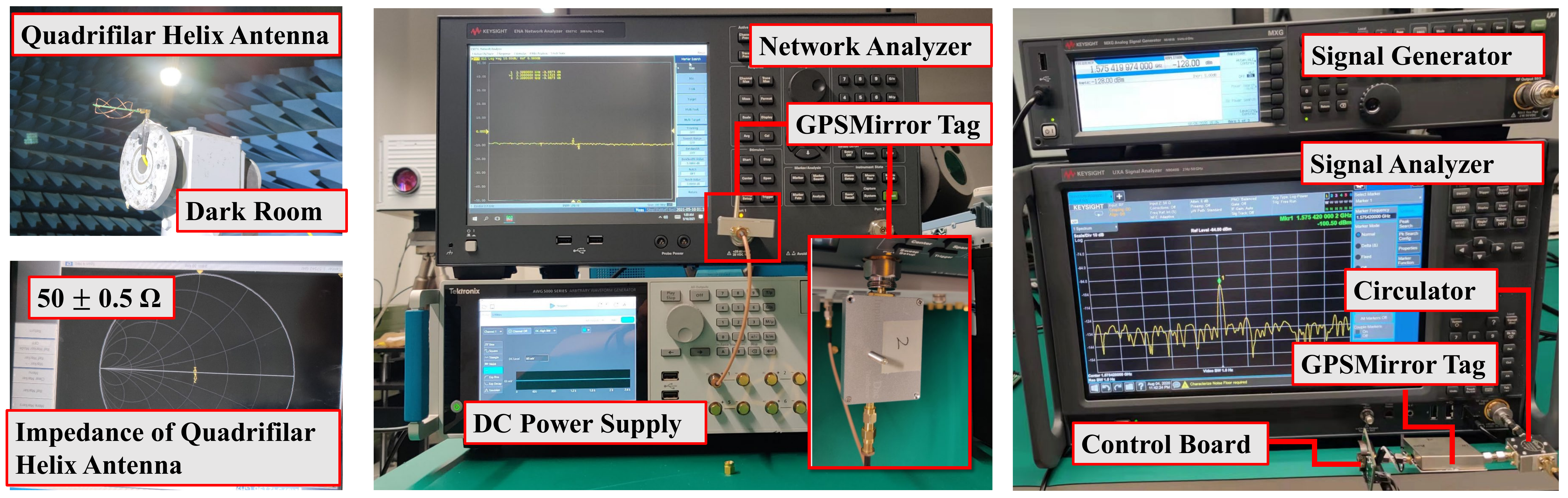}
	\caption{Hardware testbed.}
	\label{fig:hardwareTestbed} 
       \vspace{-0.6cm}
\end{figure}

\section{Evaluation} \label{sec:eval}

\begin{figure}
	\centering
	\setlength{\abovecaptionskip}{0.cm}
	\setlength{\belowcaptionskip}{0.cm}
	\subfigure[Gain with respect to voltage.]{
		\begin{minipage}[t]{0.45\linewidth}
			\centering
			\includegraphics[width=\linewidth]{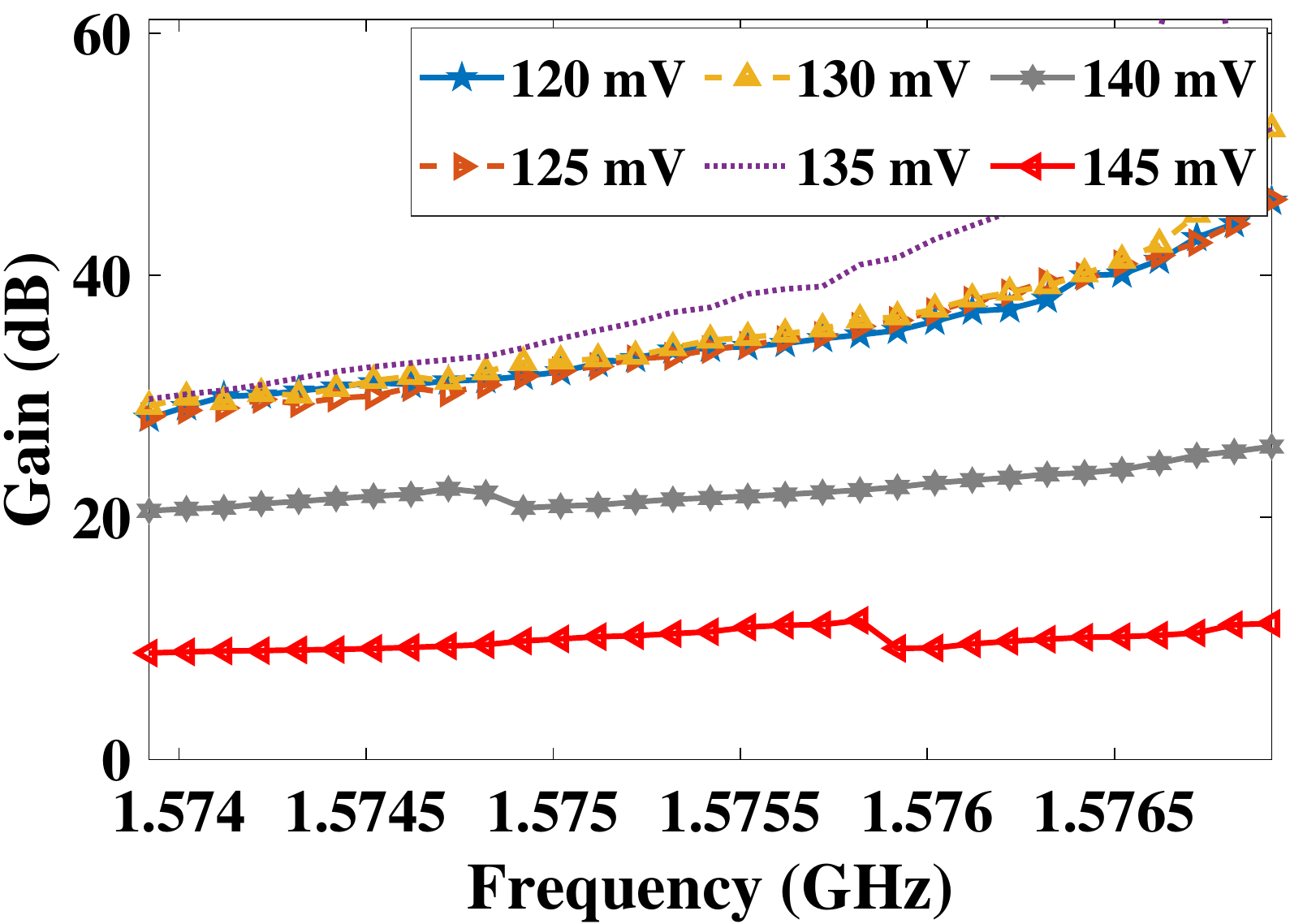}
			\label{fig:Frequence} 
		\end{minipage}
	}
	\subfigure[Gain with respect to input power.]{
		\begin{minipage}[t]{0.48\linewidth}
			\centering
			\includegraphics[width=\linewidth]{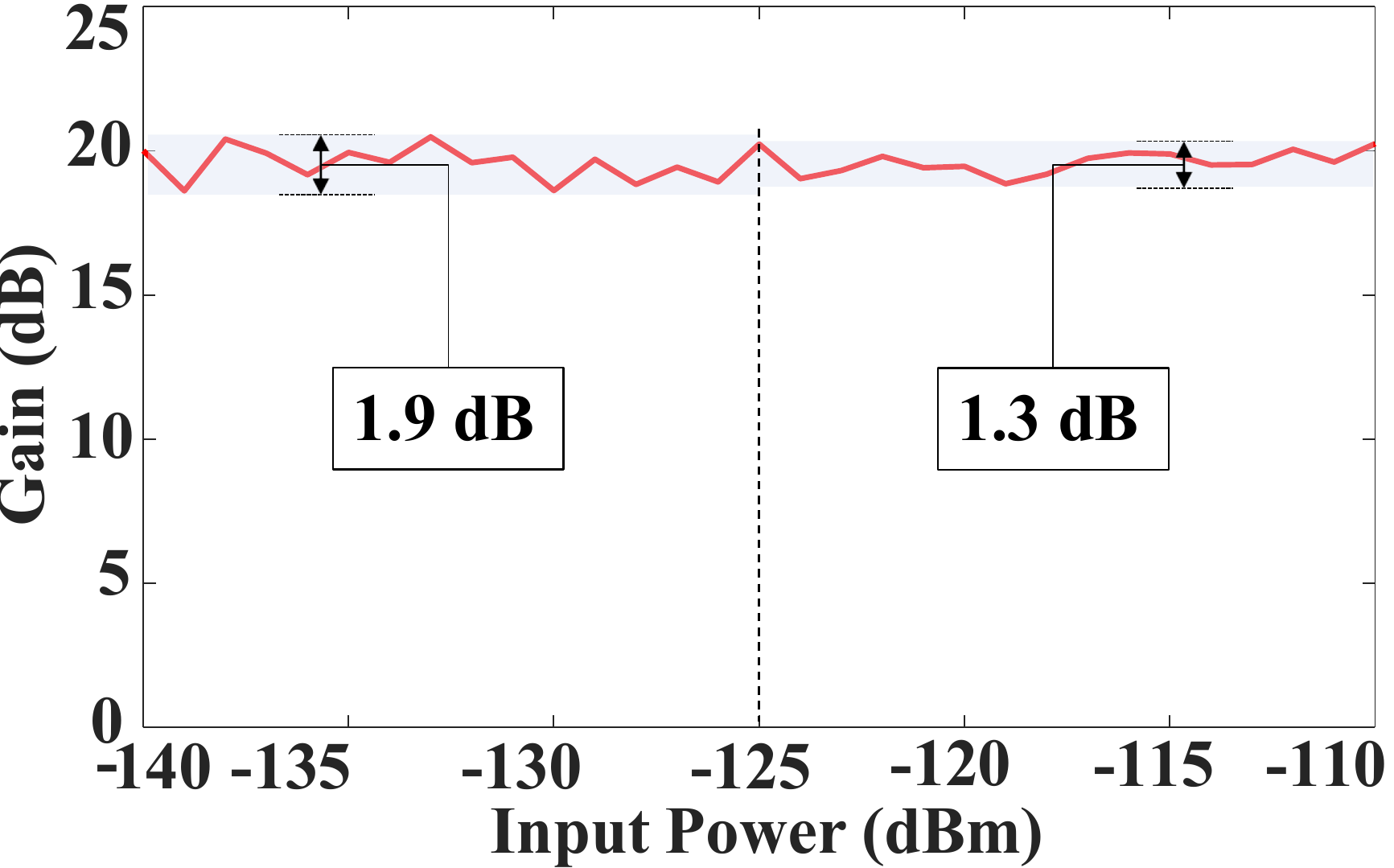}
			\label{fig:gainVsinput} 
		\end{minipage}
	}
	\caption{Hardware performance of a \sysname tag.}
	
	\vspace{-0.6cm}   
\end{figure}
In this section, we first evaluate the hardware performance of the \sysname tags with signal analyzers and smartphones. Then we conduct a series of experiments to test its coverage and localization performance under different conditions.
 \vspace{-0.3cm}
\subsection{Experimental Setup}

\textbf{Hardware testbed.} The hardware testbed is used to profile the \sysname tag, including the gains in the L1 band, the impedance of antennas, and the front-end circuit. As shown in \fig~\ref{fig:hardwareTestbed}~(b), we use the network analyzer E5061B in an electromagnetic darkroom to measure the impedance and radiation pattern of the antenna and to tune the antenna to 50~ohms in the L1 band. 
The network analyzer N5225B is used to measure the reflection coefficient of the tag with a weak excitation power. The AWG 5000 arbitrary waveform generator provides accurate bias voltages to the \sysname tags.
We use the Keysight MXG analog signal generator N5182B and the Keysight UXA signal analyzer N9020A to measure the gain of the \sysname tags. 
N5182B can generate RF signals as low as -140~dBm, and N9020A has a typical sensitivity of -150~dBm.
\begin{figure}
	\centering
	\setlength{\abovecaptionskip}{0.cm}
	\setlength{\belowcaptionskip}{0.cm}
	\subfigure[Smartphones.]{
		\begin{minipage}[t]{0.3\linewidth}
			\centering
			\includegraphics[width=\linewidth]{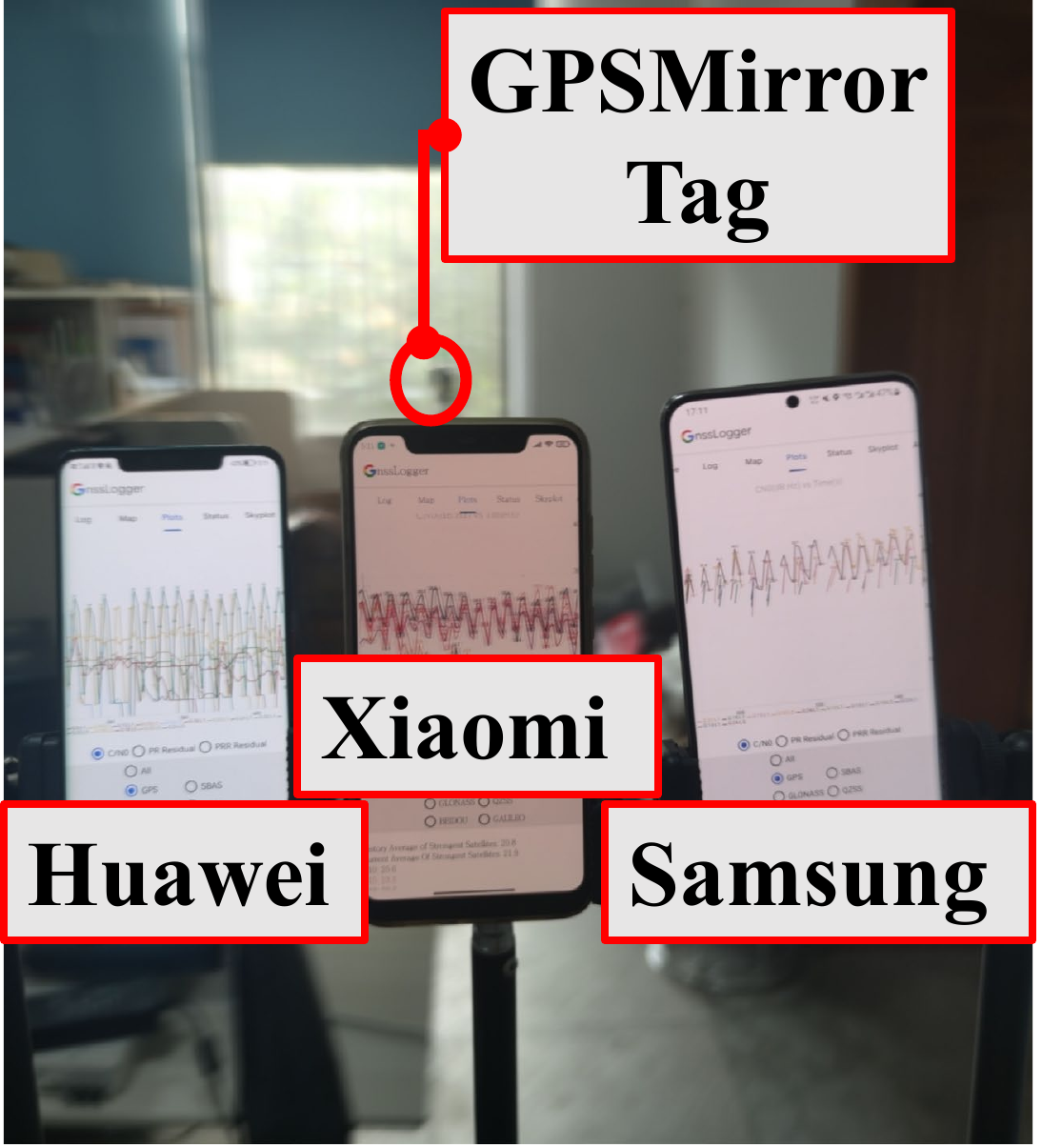}
		\end{minipage}
	}
	\subfigure[Gain detected on different smartphones with respect to distance.]{
		\begin{minipage}[t]{0.5\linewidth}
			\centering
			\includegraphics[width=\linewidth]{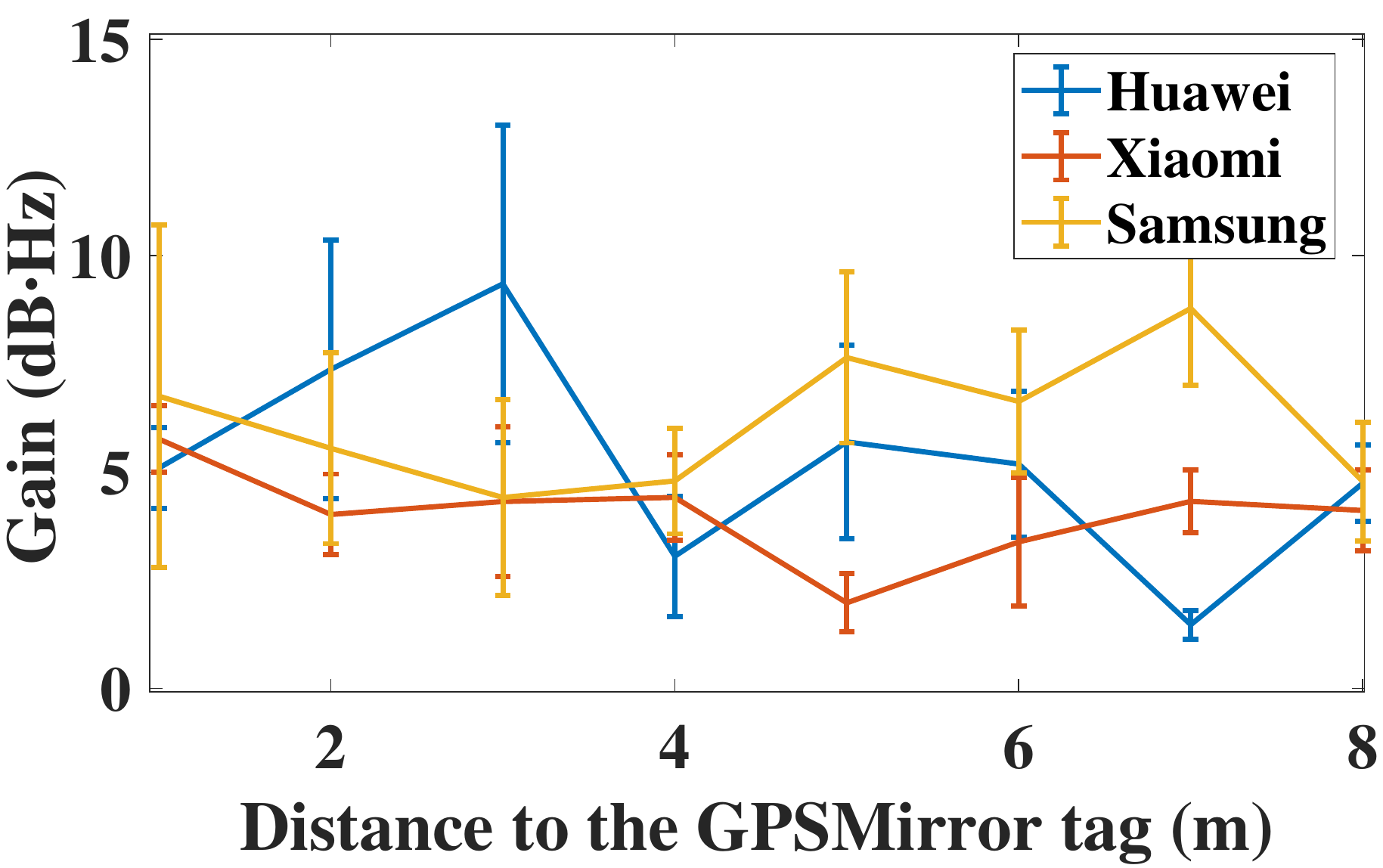}
		\end{minipage}
	}
	\caption{Reception on various smartphones.}
	\label{fig:receptionOnSmartphones} 
	\vspace{-0.7cm}   
\end{figure}

\textbf{Ground truth.} We use a Xiaomi LS-P laser range finder with a range error of 3~mm to obtain the ground truth for distance measurements.
We use a Unistrong P40 board in real-time kinematic positioning (RTK) mode to obtain the ground truth for coordinates measurements. The position error of the Unistrong P40 in RTK mode is 8~mm when synchronized with differential stations.
However, the Unistrong P40 board often fails to sync when it gets close to buildings or flyovers. We thus measure the coordinates of test points at least 30~meters away from buildings and derive all the required coordinates based on the distances measured by the laser range finder.  

\textbf{Deployment of \sysname tags.}
The \sysname tags can be deployed outside the building, in openings of a building or in unenclosed parts of flyovers to provide GPS service for pedestrians. The locations of the \sysname tags are registered in the map or floor plan and stored on smartphones.

\vspace{-0.4cm}
\subsection{GPS Signal Re-radiation}

\textbf{Performance profiling with a hardware testbed.} To measure the gain of the \sysname tag, we use the signal generator to generate a -140~dBm RF signal that passes through the \sysname tag to the spectrum analyzer. The gain is obtained by comparing the signal strength of the signal generator's output to the measured signal strength of the signal analyzer.
The results are presented in \fig~\ref{fig:Frequence}, where we can see the \sysname tag achieves about 22~dB gain with negligible distortion when the bias voltage is set at 140mV and higher gains are reachable when the bias voltage changes. We also evaluated the gain with different input signal power and the results are presented in \fig~\ref{fig:gainVsinput}, where we can see the \sysname tag achieves flat gain with distortions less than 1.3~dB when the input power higher than -125~dBm and 1.9~dB when input power higher than -140~dBm. Thus, we can conservatively conclude that the \sysname tag has sufficient sensitivity ( $\leq$-125~dBm) for scattering GPS signals.

\begin{figure}
	\centering
	\setlength{\abovecaptionskip}{0.cm}
	\setlength{\belowcaptionskip}{0.cm}
	\includegraphics[width=0.9\linewidth]{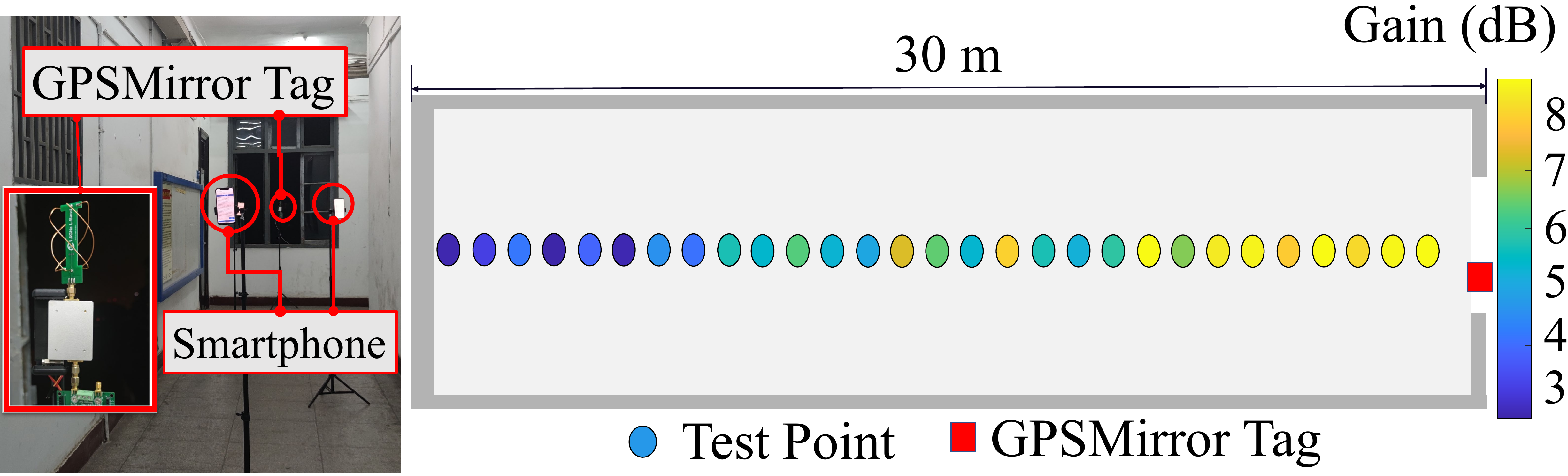}
	\caption{Coverage test in a corridor.}
	\label{fig:coverageCorridor} 
	\vspace{-0.9cm}
\end{figure}
\textbf{Received signal quality on smartphones.} We test the signal quality on three different smartphones, including Samsung Galaxy S20, Huawei Mate20 Pro and Xiaomi 8, to verify the compatibility. 
As plotted in \fig~\ref{fig:receptionOnSmartphones}, we deploy a \sysname tag in close to a window and simultaneously collect data using three smartphones. We compute the statistical average gain, along with a 95\% confidence interval, for distances ranging from 1~m to 8~m away from the tag. 
Results demonstrate that smartphones from a variety of manufacturers are capable of effectively receiving GPS signals scattered by the tags.

\vspace{-0.3cm}
\subsection{Coverage of \sysname Tags}\label{sec:evalCoverage}

\begin{figure*}
	\hfill
	\begin{minipage}{0.65\linewidth}  
		\setlength{\abovecaptionskip}{-0.1cm}
		\begin{minipage}[t]{0.49\linewidth}
			\raggedleft
			\subfigure[Tripod-Mounted]{
				\includegraphics[width=\linewidth]{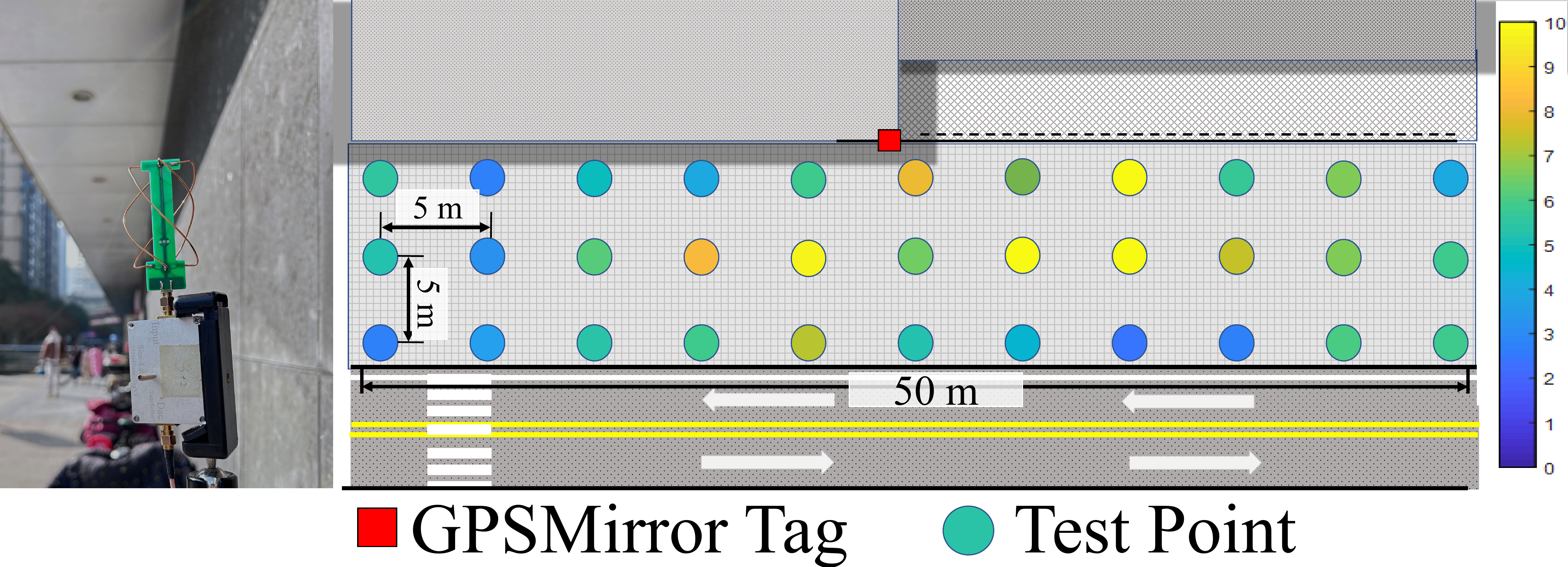}
				\label{fig:urbanTripodMounted}
			}
		\end{minipage}
		\hfill
		\begin{minipage}[t]{0.49\linewidth}
			\raggedleft
			\subfigure[Wall-Mounted]{
				\includegraphics[width=\linewidth]{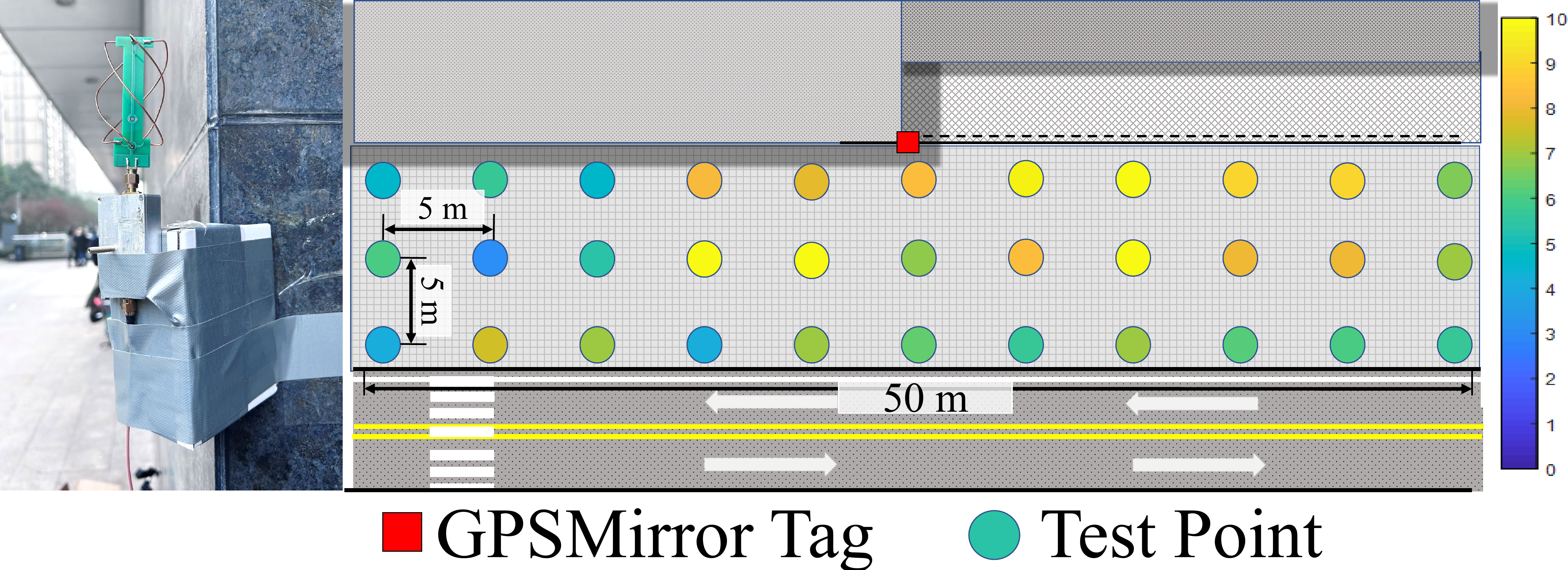}
				\label{fig:urbanWallMounted}
			}
		\end{minipage}
		\caption{Coverage performance of a \sysname Tag in urban canyon.}
		\label{fig:coverageUrban}
	\end{minipage}
	\begin{minipage}{0.32\linewidth}  
		\raggedright
		\setlength{\abovecaptionskip}{0.1cm}
		\includegraphics[width=\linewidth]{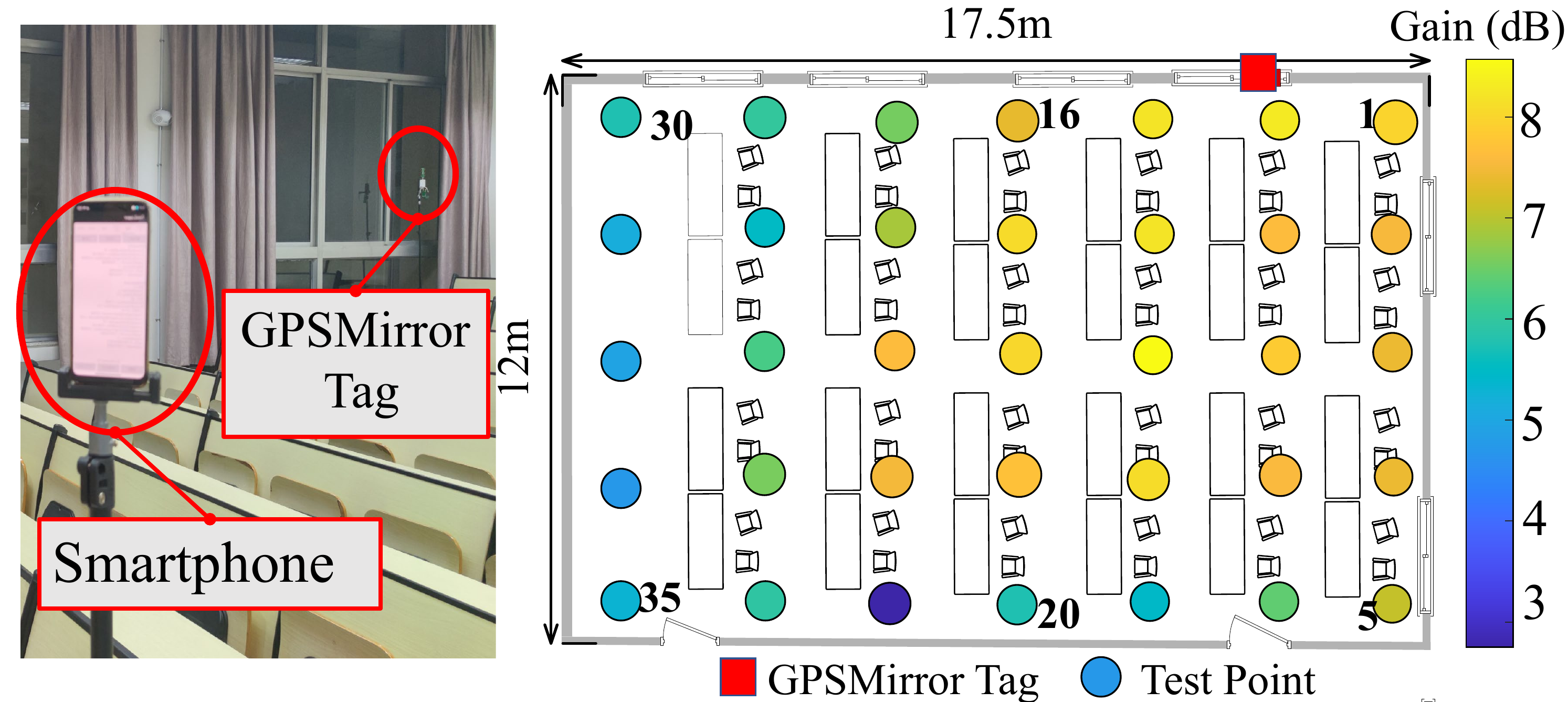}
		\caption{Coverage test in a room.}
		\label{fig:coverageIndoors}
	\end{minipage}
\end{figure*}

\begin{figure*}
	\hfill
	\setlength{\abovecaptionskip}{0cm}
	\begin{minipage}[t]{0.32\linewidth}
		\centering
		\subfigure[Tripod-Mounted]{
			\includegraphics[width=\linewidth]{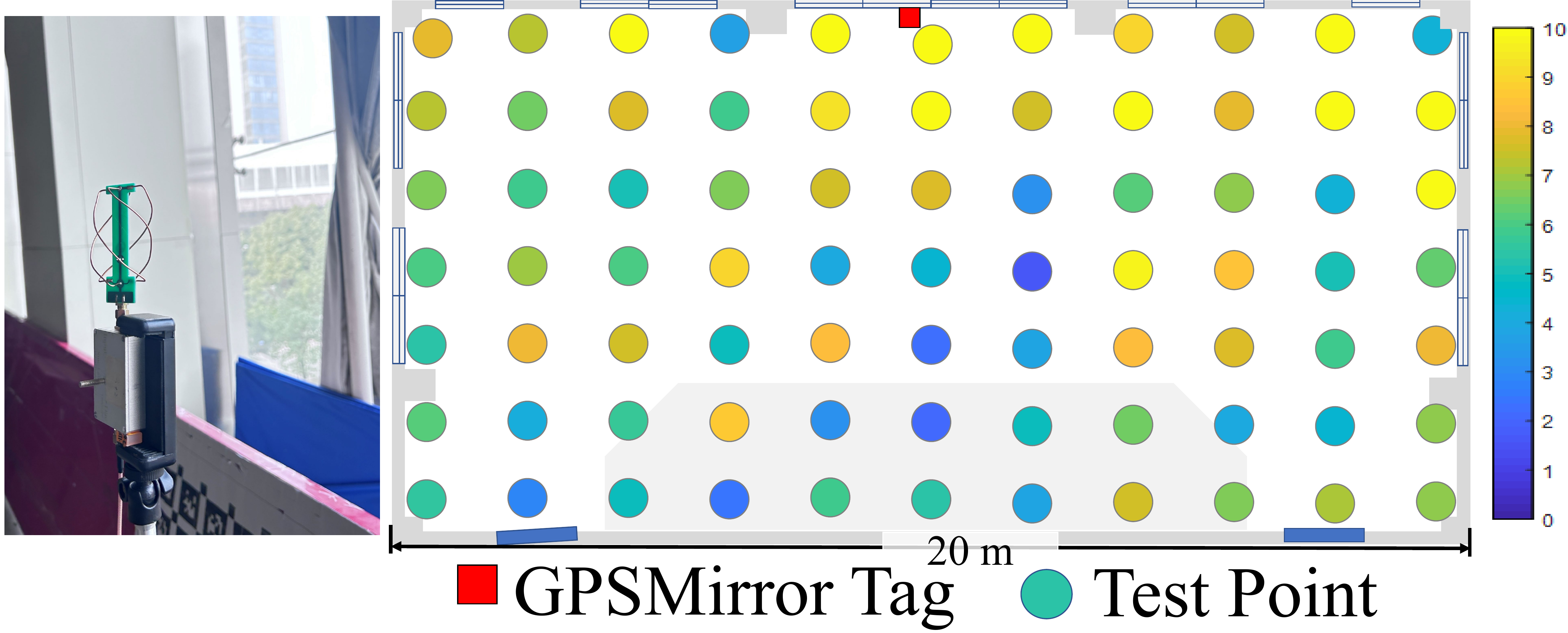}
			\label{fig:roomTripodMounted}
		}
	\end{minipage}
	\hfill
	\begin{minipage}[t]{0.32\linewidth}
		\centering
		\subfigure[Window-Mounted]{
			\includegraphics[width=\linewidth]{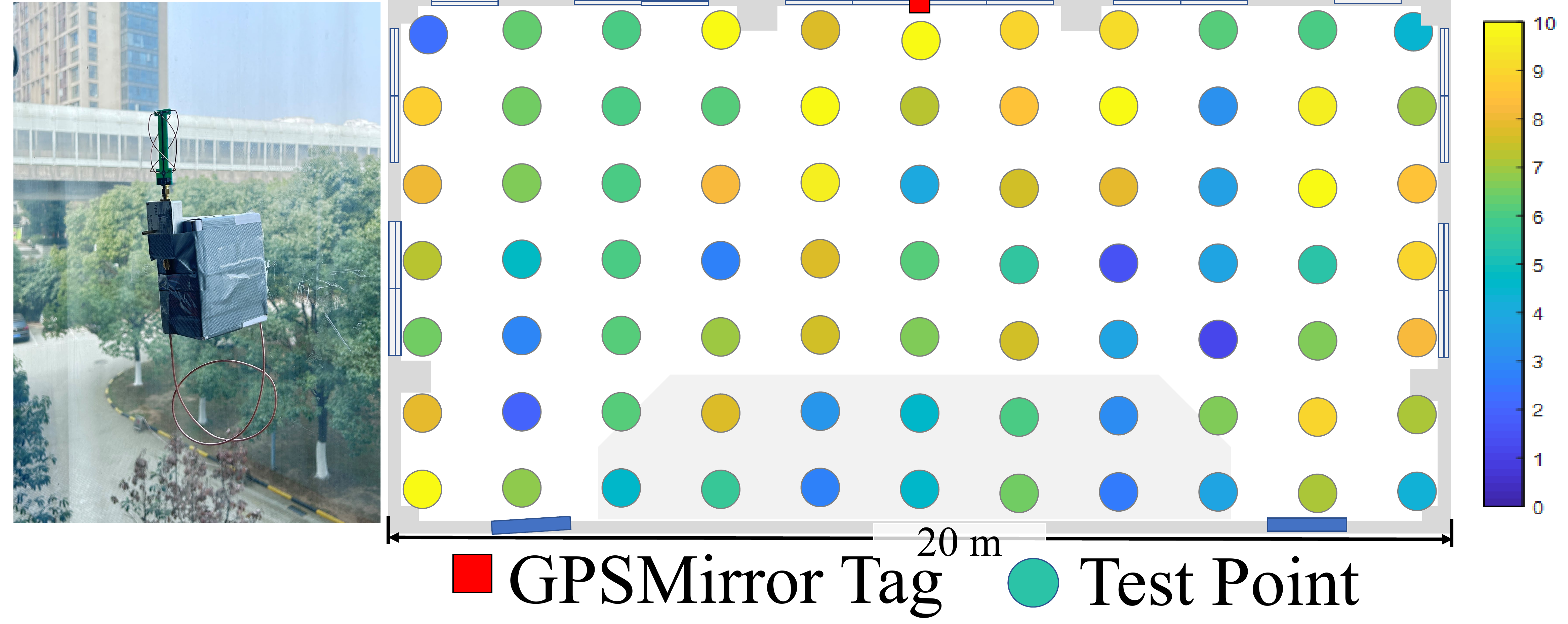}
			\label{fig:roomWindowMounted}
		}
	\end{minipage}
	\hfill
	\begin{minipage}[t]{0.34\linewidth}
		\centering
		\subfigure[Wall-Mounted]{
			\includegraphics[width=\linewidth]{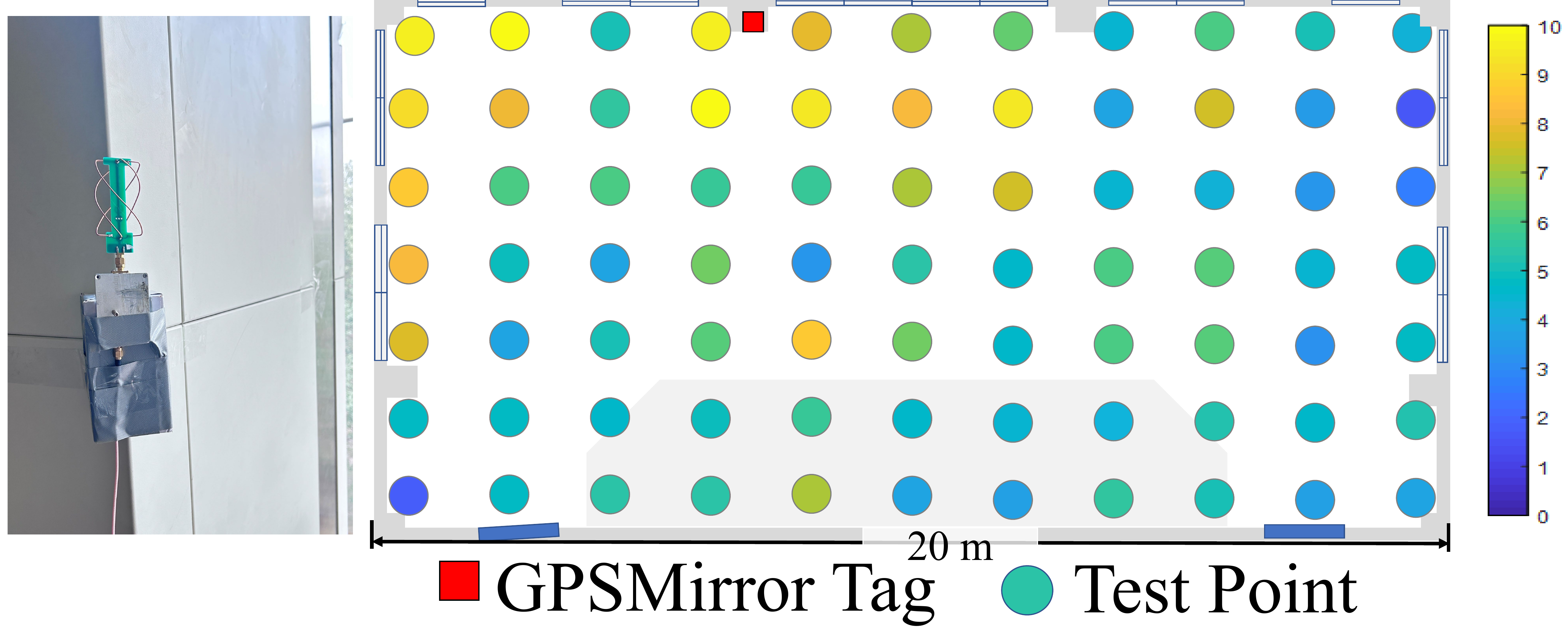}
			\label{fig:roomWallMounted} 
		} 
	\end{minipage}
	\caption{Coverage performance of a \sysname tag with different mounting methods in a flat room.}
	\label{fig:coverageLecture} \vspace{-0.5cm}
\end{figure*}
We define the coverage as the region where a \sysname tag's ``ON-OFF'' switching can be detected and smartphones can position with the tag.
\sysname enhances the GPS coverage in two main aspects. On the one hand, \sysname tags alter the propagation paths to increase the number of visible satellites. On the other hand, the GPS signals originally reachable to GPS receivers are strengthened by the \sysname tags. To evaluate the coverage performance from both aspects, we deploy \sysname tags in various scenarios including a 30m-long corridor, a $50\times12~m^2$ sidewalk in an urban canyon, a $17.5\times12~m^2$ classroom and a $20\times14~m^2$ lecture hall. We set test points in each scenario and use smartphones to record raw $C/N_0$ measurements of both scattered and non-scattered GPS signals for 5 minutes at each test point. We use the $C/N_0$ of scattered GPS signals to minus that of non-scattered signals to present the enhanced performance. Then, we plot the mean value of 5 minutes in each test point as shown from \fig~\ref{fig:coverageCorridor} to \fig~\ref{fig:coverageLecture}.   

In all scenarios, \sysname tags strengthen the GPS signal by at least 3~dB including the farthest location. Specifically, at the farthest test point in \fig~\ref{fig:coverageUrban}, which is 27.7~m away from the tag, there is >4~dB gain from the tag in both wall-mounted and tripod-mounted conditions. Since the nearest test point is about 2~m to the tag with gains of 9~dB, the drop-off of the gain is about 5~dB in this condition. Moreover, the farthest test point in \fig~\ref{fig:coverageLecture} is about 15.6~m to the tag with >3~dB in the three conditions. Therefore, we can conclude that the drop-off of gain is slow.  
Besides, \fig~\ref{fig:coverageCorridor} shows that the \sysname tag can enable sufficient coverage within a radius of up to 30~m. Thus, the coverage of \sysname tags reaches the limit of the US regulation~\cite{NTIA_RedBook}, i.e., 30~m.

\textbf{Impact of mounting methods.}
To evaluate the impact of different mounting methods, we conduct experiments to evaluate the performance of wall-mounted, window-mounted and tripod-mounted tags in both indoor and outdoor scenarios.
Specifically, \fig~\ref{fig:coverageUrban} shows the coverage performance of a \sysname tag with different mounting methods in an urban canyon. We set 33 test points and each point is spaced 5~m apart. \fig~\ref{fig:coverageLecture} shows the coverage of \sysname tags in a lecture hall with different mounting methods. We set 77 test points and each one is spaced 2~m apart. Results show that different mounting methods have a negligible impact on the coverage.

\textbf{Impact of deployment location.} We notice that the deployment location of the \sysname tag influences the gain's distribution in these test points. We compare \fig~\ref{fig:roomTripodMounted},\ref{fig:roomWindowMounted} and \fig~\ref{fig:roomWallMounted} can draw a conclusion that it is better to deploy the \sysname tag close to the center of the open area for more uniform signal enhancement.

\vspace{-0.7cm}

\subsection{Positioning Accuracy} \label{Sec:PosAccurate}
To fully exploit the performance of the positioning accuracy, we conduct the following experiment to evaluate the performance of both static and dynamic positioning. Further, we conduct a controlled experiment to explore the potential of using multiple \sysname tags for high positioning accuracy. 

\begin{figure}   
	\centering
	\setlength{\abovecaptionskip}{3pt}
	\setlength{\belowdisplayskip}{3pt}
	\includegraphics[width=0.75\linewidth]{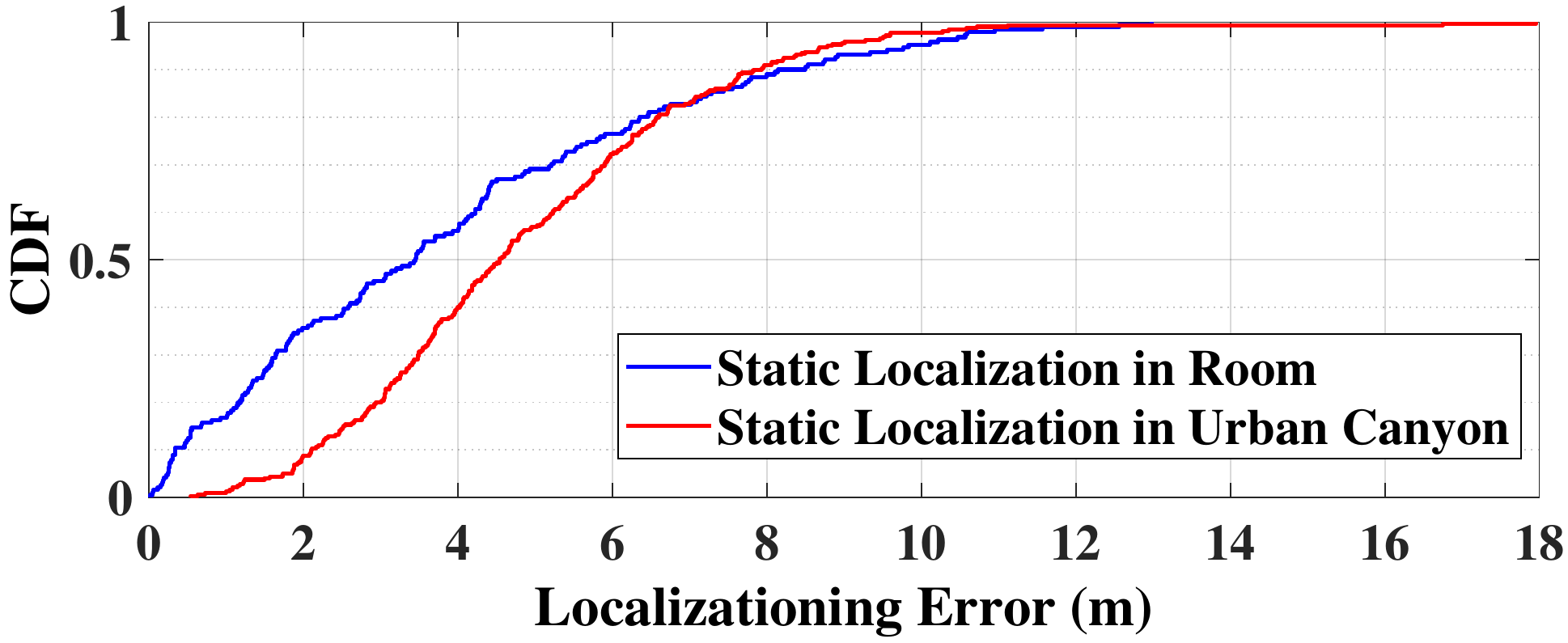}
	\caption{Positioning accuracy.}
	\label{fig:positioningAccuracy} 
	\vspace{-0.4cm}
\end{figure}

\textbf{Static positioning.} We evaluate the static positioning accuracy of the \sysname system in the following two scenarios: indoors, as shown in \fig~\ref{fig:coverageIndoors} and \ref{fig:coverageLecture}, and in an urban canyon, as shown in \fig~\ref{fig:coverageUrban}. To assess performance, we plot the cumulative distribution function (CDF) using the average position error of each point. The median positioning error is about 3.7~m in the indoor scenario and 4.6~m in the urban canyon as plotted in \fig~\ref{fig:positioningAccuracy}.
The localization accuracy in the indoor scenario is a little better than that in urban canyons. We speculate that this effect may be caused by the following reasons. The propagation path through a \sysname tag in an indoor scenario is more clear since windows are the only opening to the sky. On the contrary, GPS signals in an urban canyon are more complex since smartphones may receive more signals from propagation paths other than passing through the \sysname tags. Overall, the accuracy in both scenarios improves a lot with a \sysname tag eliminating the common delay errors. 

\begin{figure}   
	\centering
	\setlength{\abovecaptionskip}{0.cm}
	\begin{minipage}[t]{0.55\linewidth}
		\centering
		\includegraphics[width=\linewidth]{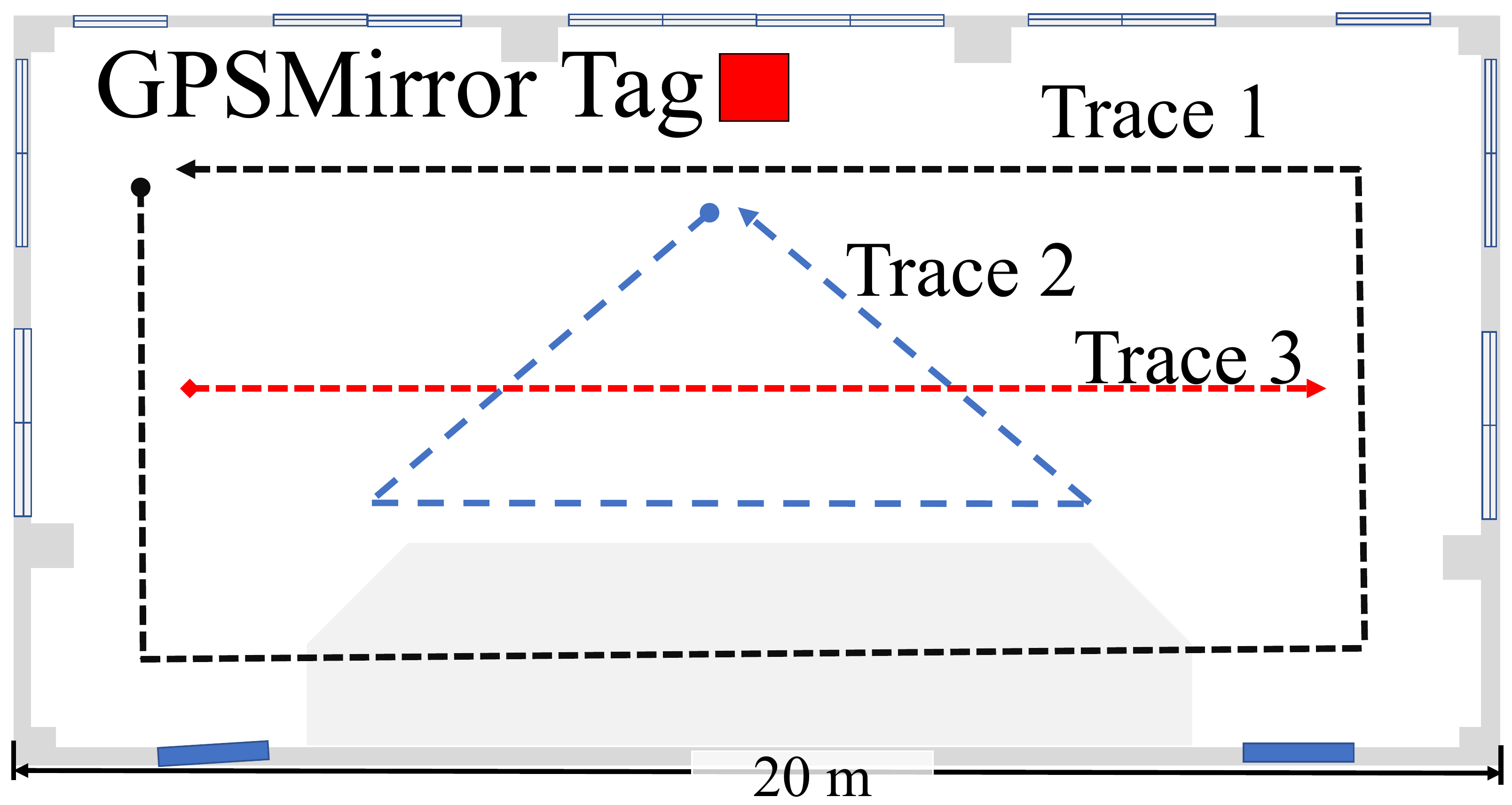}
	\end{minipage}
	\begin{minipage}[t]{0.37\linewidth}
		\centering
		\includegraphics[width=\linewidth]{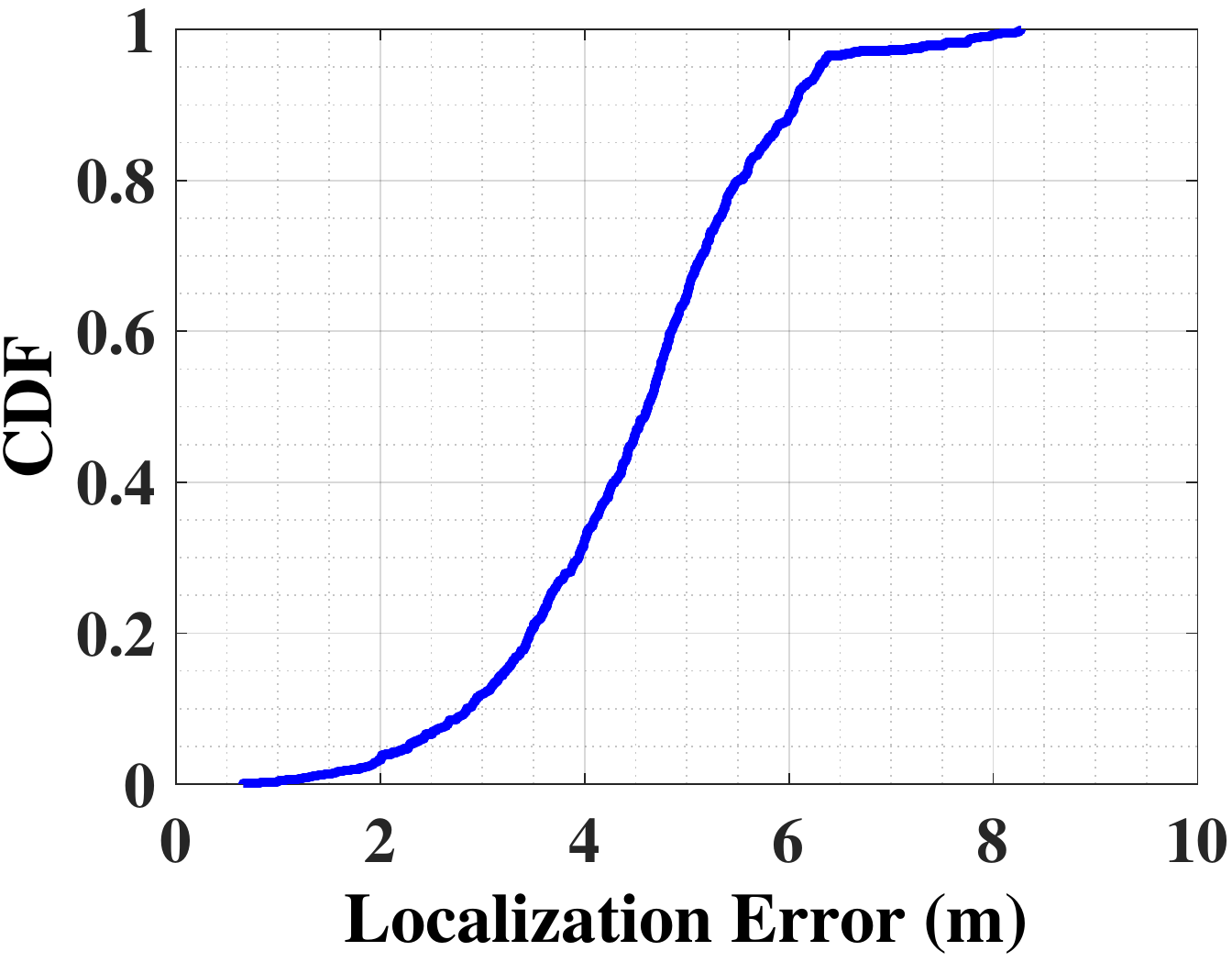}
	\end{minipage}
	\caption{Dynamic localization with a \sysname tag.}
	\label{fig:dynamicLocalization} 
	\vspace{-0.5cm}
\end{figure}

\begin{figure*}[t]
	\hfill
	\begin{minipage}{0.64\linewidth}  
		\setlength{\abovecaptionskip}{0cm}
		\begin{minipage}[t]{0.6\linewidth}
			\raggedleft
			\includegraphics[width=\linewidth]{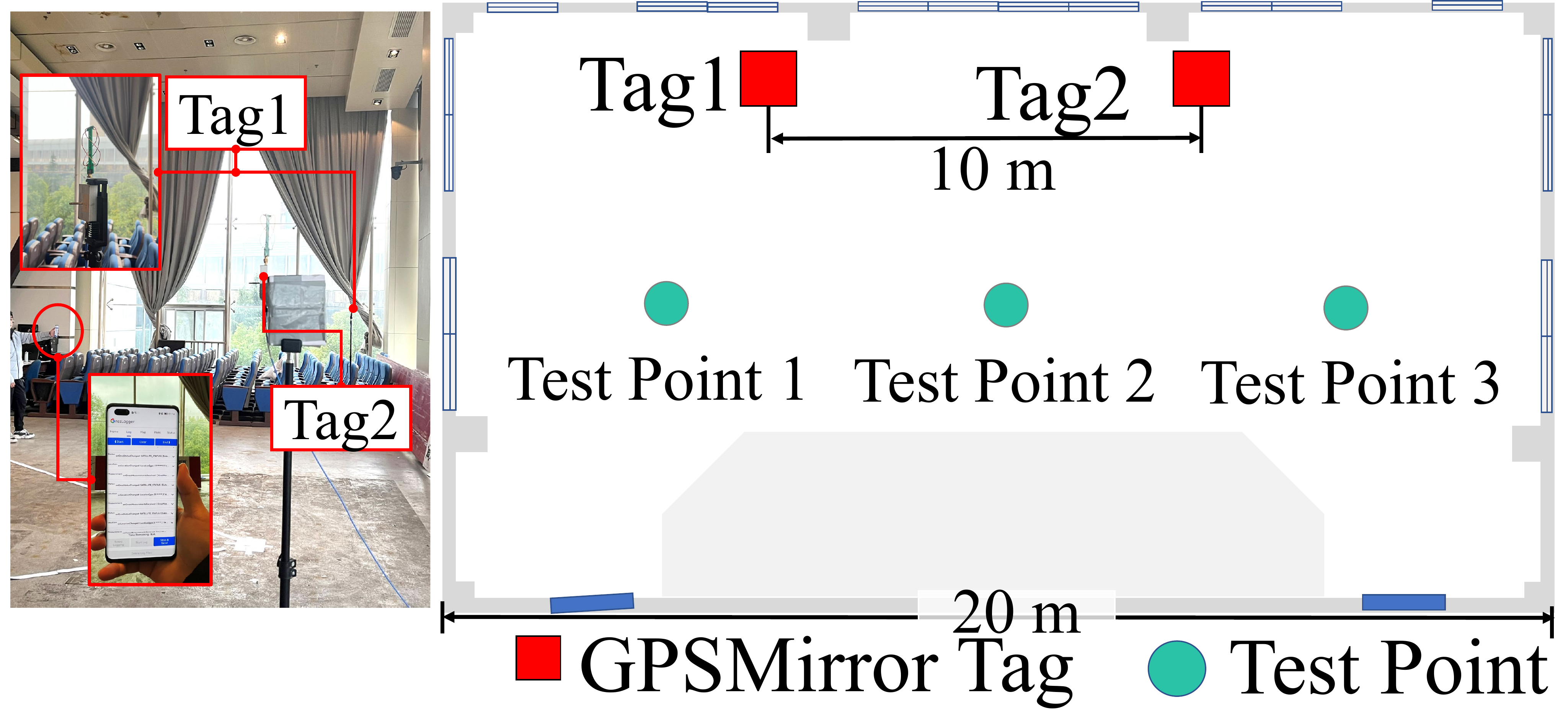}
		\end{minipage}
		\hfill
		\begin{minipage}[t]{0.37\linewidth}
			\includegraphics[width=\linewidth]{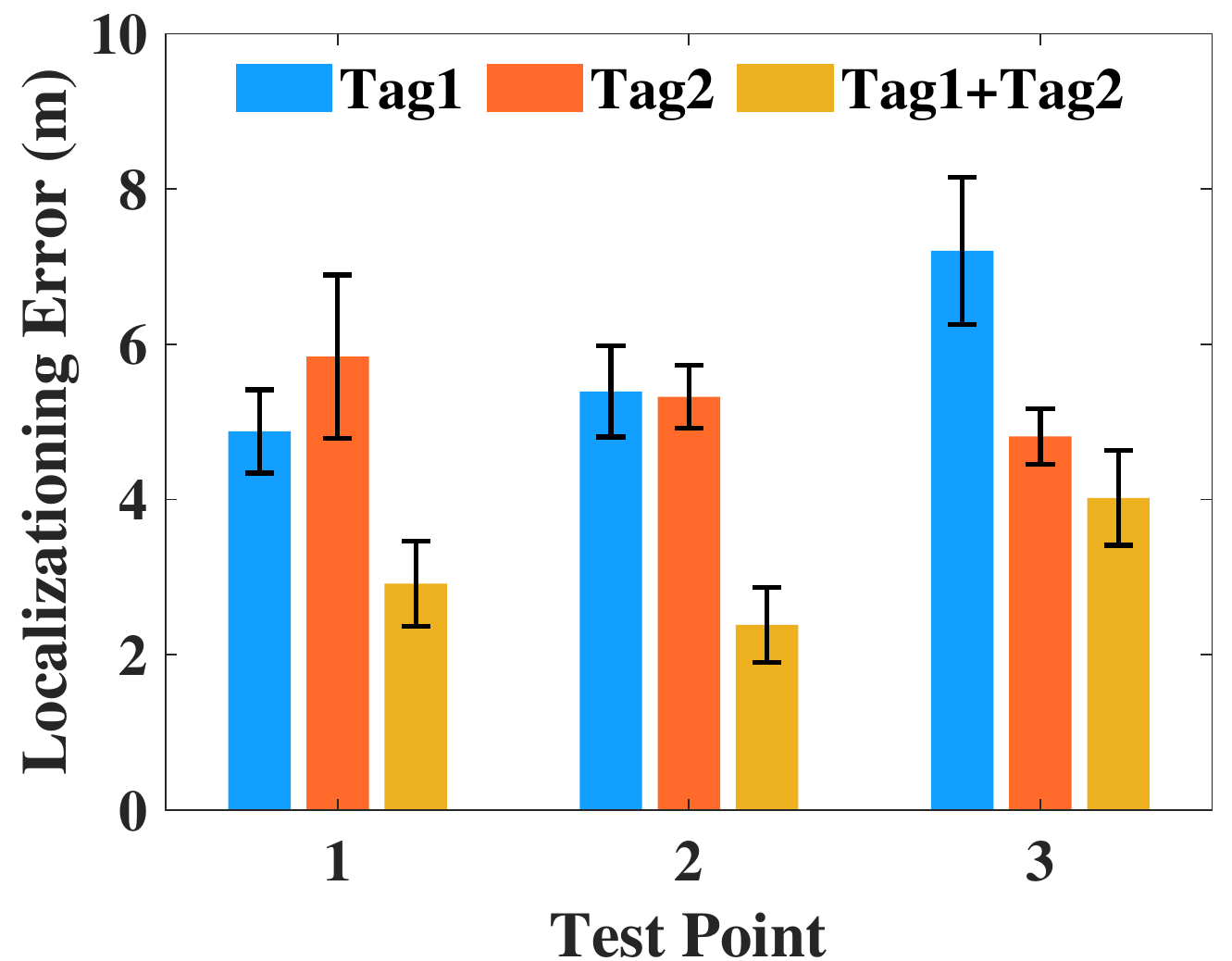}
		\end{minipage}
		\caption{Positioning performance with multiple tags.}
		\label{fig:positioningAccuracyMultiTag}
	\end{minipage}
	\begin{minipage}{0.34\linewidth}  
		\centering
		\setlength{\abovecaptionskip}{0.cm}
		\includegraphics[width=\linewidth]{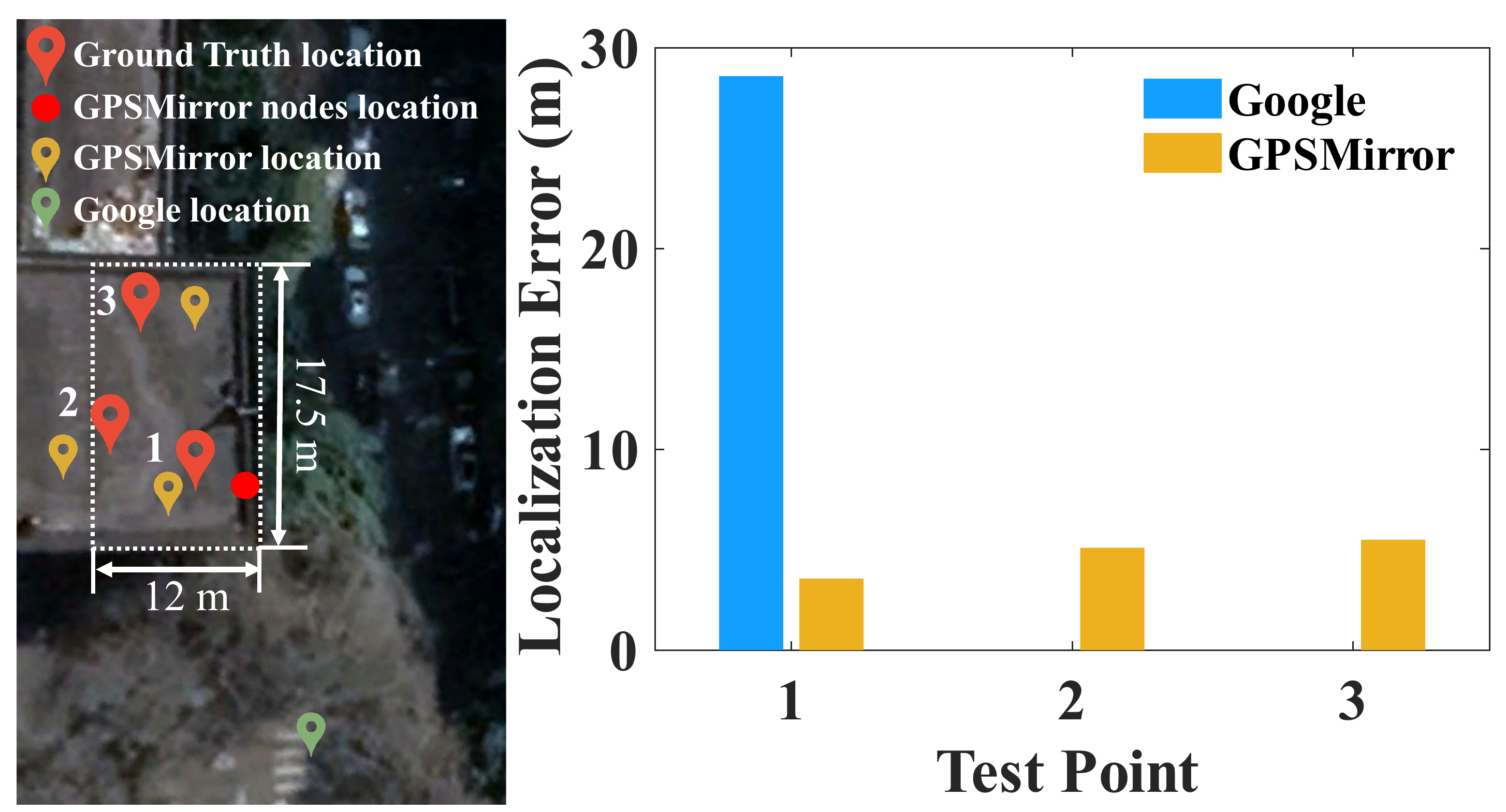}
		\caption{Large flat room.}
		\label{fig:accuracyIndoors}
	\end{minipage}  
	\vspace{-0.4cm}
\end{figure*}
\begin{figure*}[h]
	\hfill
	\begin{minipage}[t]{0.3\linewidth}
		\centering
		\setlength{\abovecaptionskip}{0.cm}
		\includegraphics[width=\linewidth]{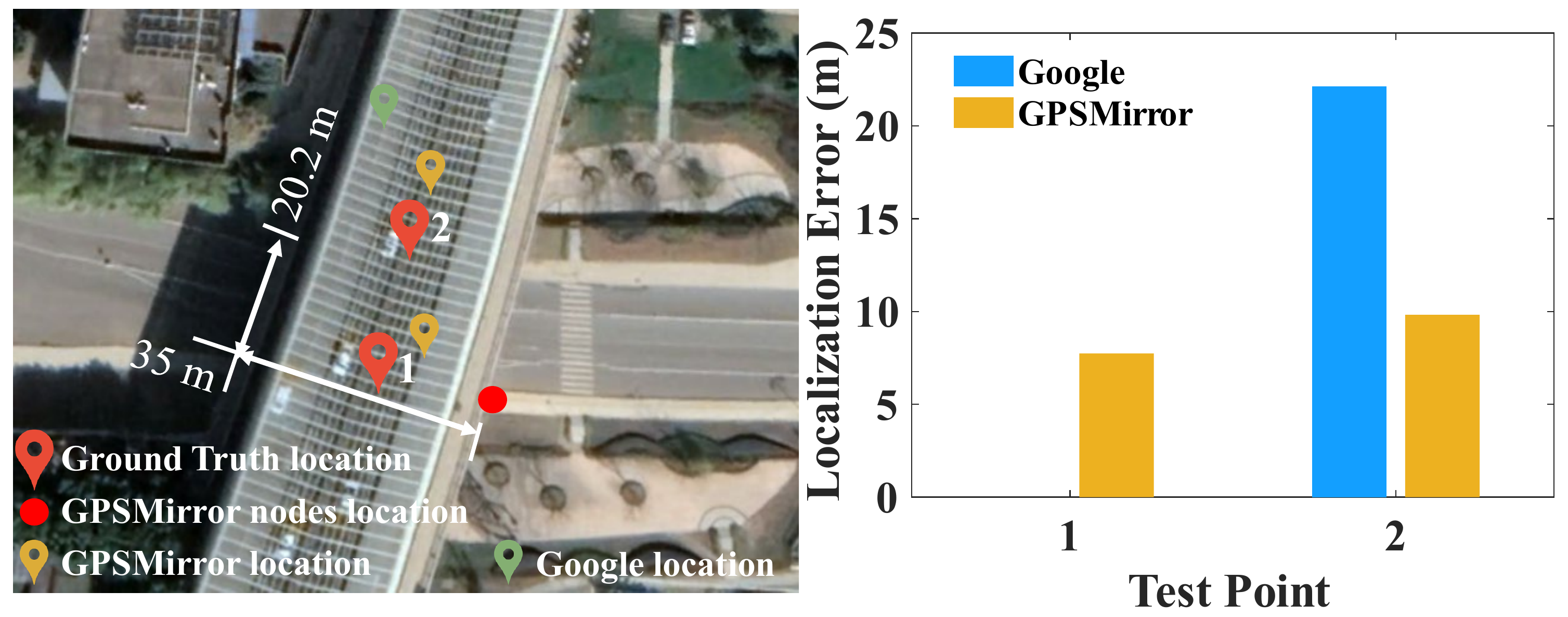}
		\caption{Under flyover.}
		\label{fig:accuracyFlyover}
	\end{minipage}
	\hfill
	\begin{minipage}[t]{0.33\linewidth}
		\centering
		\setlength{\abovecaptionskip}{0.cm}
		\includegraphics[width=\linewidth]{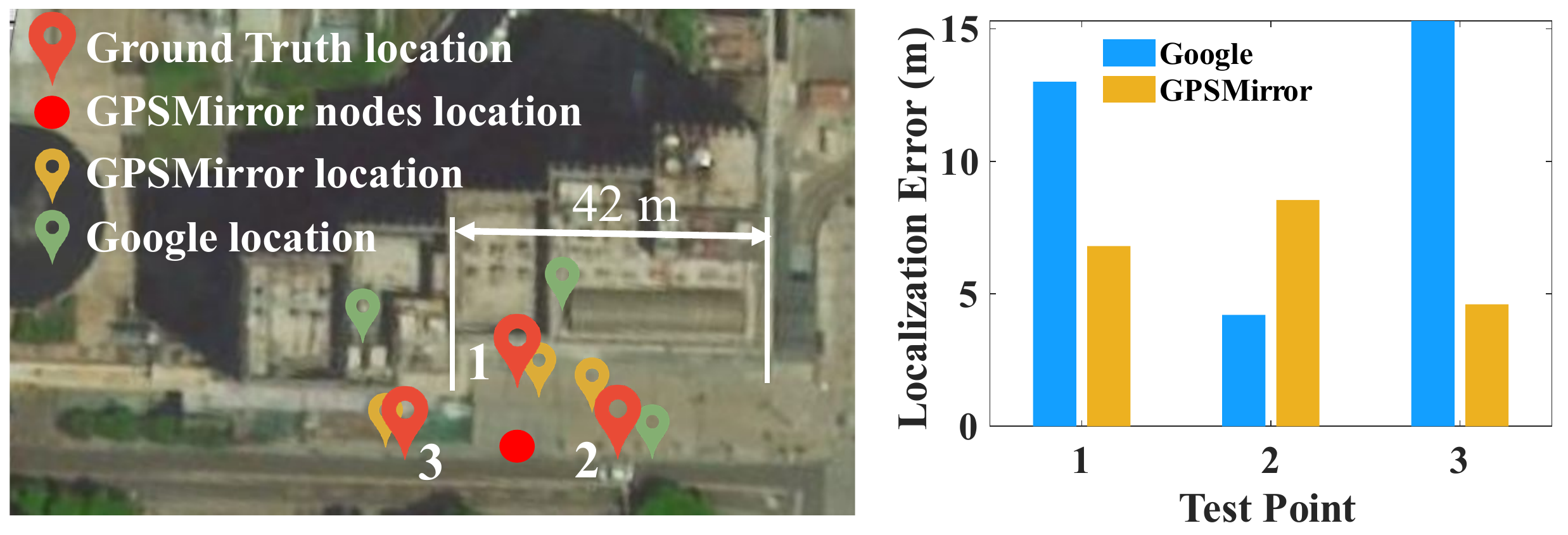}
		\caption{Exterior of buildings.}
		\label{fig:accuracyUrban}
	\end{minipage}
	\hfill
	\begin{minipage}[t]{0.31\linewidth}
		\centering
		\setlength{\abovecaptionskip}{0.cm}
	      \setlength{\belowdisplayskip}{3pt}
		\includegraphics[width=\linewidth]{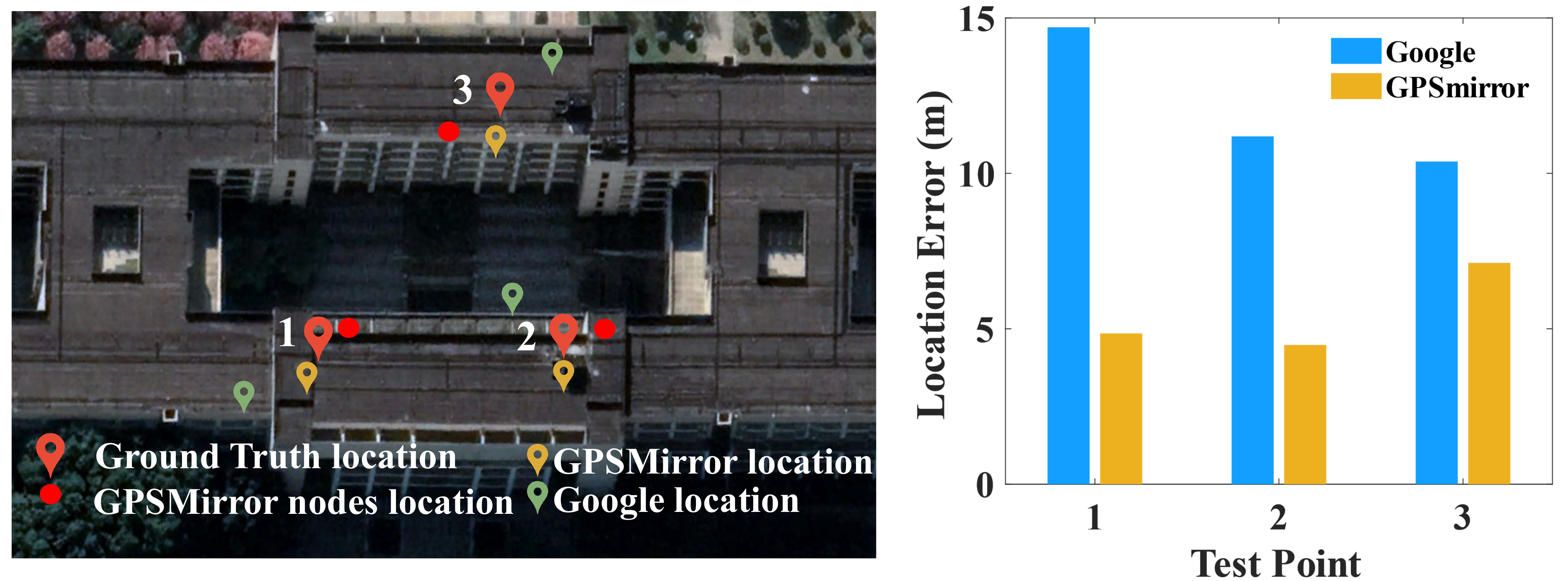}
		\caption{Center patio.}
		\label{fig:accuracyCenterPatio}
	\end{minipage}
	\vspace{-0.4cm}
\end{figure*} 

\textbf{Dynamic positioning.} To evaluate the performance of dynamic positioning, we perform a controlled experiment with a \sysname tag mounted on the window and set 3 types of trajectories for the receiver as shown in \fig~\ref{fig:dynamicLocalization}. A volunteer walks along the trajectories with an average speed of about 25~cm/s per second and holds a smartphone recording GPS measurements. We compute the location based on every four samples and compare it with the ground truth, which is distributed every 1~m on the trajectories. The results show that we can achieve a median localization error of about 4.5~m with GPS signals only, which is sufficient in urban canyons and some indoor regions. However, the accuracy gets a little worse than in static positioning conditions. 
We speculate this may be primarily limited by the low sampling rate of the smartphone, which is defaulted to 1~Hz for energy saving. To eliminate the common error correctly, multiple samples are needed, but the correlation between these samples is reduced more in dynamic localization conditions, resulting in slightly worse accuracy. Increasing the sampling rate can mitigate this limitation. Some GNSS sensors for smartphones support higher sampling rates, e.g., 40Hz~\cite{UbloxD9PMP}. It is possible to achieve this rate through firmware modifications~\cite{shade2018android}. 

\textbf{Positioning with multiple tags.} 
To investigate the potential for enhancing localization accuracy using multiple \sysname tags, we conduct a controlled experiment with two tripod-mounted tags offering GPS-enhancing services simultaneously. These tags are synchronized via cable and operated on a time-division schedule. We established three test points and collected 180 samples at each using a handheld smartphone. The smartphone's position was estimated with Tag1 and Tag2, and both base vectors from the tags were used for estimation. As illustrated in \fig~\ref{fig:positioningAccuracyMultiTag}, the average positioning accuracy of two tags at the three test points was approximately 2.6~m, 2.4~m, and 3.6~m, with 95$\%$ confidence interval error bars. Localization results using two tags are significantly better than those using a single tag.

\vspace{-0.3cm}
\subsection{Case Studies}
We deploy the \sysname tags and evaluate the positioning accuracy of our system in four real-world scenarios, i.e., large flat room (Factory, \fig~\ref{fig:accuracyIndoors}), flyover (\fig~\ref{fig:accuracyFlyover}), exterior of tall buildings (Urban canyon, \fig~\ref{fig:accuracyUrban})), and center patio (Urban canyon, \fig~\ref{fig:accuracyCenterPatio}). We use an open-source GPS position estimation project provided by Google as baseline~\cite{googleGNSS}. 

\fig~\ref{fig:accuracyIndoors} shows the results in a large flat room on the ground, a common architectural structure as in many factories. We deploy a single \sysname tag in the room and evaluate the localization performance in three positions. Google's baseline algorithm fails to position in test point 2 and 3 due to insufficient visible satellites. In contrast, \sysname achieves the positioning accuracy of under 10~m for all test points. 

\fig~\ref{fig:accuracyFlyover} shows the results under a flyover, which is challenging for a single \sysname tag to provide GPS service. Because the middle is the vehicle way, the \sysname tag can only be deployed at one side, which is not the area that provides the best coverage. We evaluate the positioning performance in two test points. As expected, the positioning accuracy decreases on the opposite side of the road.

The exterior of tall buildings is a common view in urban areas. We deploy a single \sysname tag 12~m away from the building and try to provide high-accuracy GPS service near the building. In position 1 and position 3, \sysname achieves average performance while Google's baseline algorithm degrades rapidly. However, the positioning accuracy of Google's baseline in position 2 (LoS region) achieves better performance. This is as expected because the position is far from the \sysname tag while conventional GPS positioning algorithms can achieve good performance in LoS regions, which are not our target cases.

The center patio is a common architectural structure in an urban area, which is surrounded by tall buildings with only the middle part having a view of the sky. We deploy \sysname tags here to further evaluate the generalizability of our tags in downtown mega-cities. As shown in \fig~\ref{fig:accuracyCenterPatio}, \sysname significantly improved the position accuracy for all test points.

\vspace{-0.3cm}

\section{Related Work}\label{sec:related}
\textbf{GPS relays.}
The commonly used GPS relays~\cite{GPSrelayGRK, GPSrelayQGL1, GPSrelayRGA30-DV} mainly consist of two parts. The first part includes an antenna and an LNA
deployed outdoors, to receive the signals from GPS satellites. The second part is deployed inside the building and consists of a GPS transmit antenna and a power amplifier. These two parts are connected by an RF cable and powered by a plug-in source. Such GPS relays are power hungry, typically consuming $W$-level power for coverage of less than 30~m~\cite{NTIA_RedBook}. They are not easy to deploy due to the need for a stable power supply and the complexity of routing RF cables. In contrast, \sysname re-radiates GPS signals by integrating an antenna and a reflection amplifier into a single module. It only consumes $\mu W$-level power and can be easily powered by a coin battery or an energy harvesting module.

\textbf{GPS and wireless positioning.}
Recent efforts~\cite{nirjon2014coin,liu2015co,liu2018gnome} have attempted to improve the performance of GPS positioning. CO-GPS~\cite{liu2015co} designs a cloud-offloaded GPS solution to save energy for portable sensing devices. GNOME~\cite{liu2018gnome} leverages Google Street View for accurate GPS positioning in urban canyons. COIN-GPS~\cite{nirjon2014coin} develops a robust GPS receiver with a purpose-built directional antenna for indoor GPS positioning. \sysname is inspired by this literature while aiming to provide practical indoor GPS positioning services for off-the-shelf smartphones. DGPS~\cite{weng2020new,morgan1995differential} uses one or more reference stations to eliminate the propagation errors for accurate positioning. \sysname shares similar principles to reduce positioning errors, but differs from DGPS in that the positioning algorithm runs entirely on unmodified smartphones, merely using the GPS application programming interfaces (APIs).

Wireless indoor positioning systems, such as Bluetooth low energy (BLE)~\cite{BLEAccuracy, CC2640R2F, Trax20220}, ultra-wide-band (UWB)~\cite{UWBAccuracyPower, UWBAccuracy, DW1000, UWBGateWay_EHIGHEH100602D04}, and WiFi~\cite{WiFiRTTAccuracy, CC3230S}, have demonstrated high precision for indoor localization. However, such specialized infrastructures entail laborious mapping and calibration at the deployment phase and also require a power supply for continuous positioning. They are not compatible with the GPS service on smartphones and have not seen wide deployment.

\textbf{Tunnel diode-based backscatter.} Recent research demonstrated that tunnel diodes enable orders of magnitude improvement in backscatter communication range with $\mu W$-level power consumption~\cite{amato2018tunnel,gumber2020nonlinear,varshney2019tunnelscatter,amato2019harmonic,varshney2020tunnel,varshney2022judo}. \sysname also uses a tunnel diode to amplify incident signals, but differs from the prior work through new noise suppression and precision impedance control mechanisms. Specifically, \sysname needs to be highly sensitive in acquiring weak GPS signals and to ensure that the scattered signals still preserve the modulated waveforms of incident signals. The backscatter design in \sysname is the first to meet these requirements. Moreover, \sysname strictly works as a reflective amplifier rather than a tunnel diode oscillator \cite{amato2018tunnel,gumber2020nonlinear,varshney2019tunnelscatter,amato2019harmonic,varshney2020tunnel}. \sysname does not need to generate a carrier by itself because it only captures and amplifies the GPS signals. Jodu~\cite{varshney2022judo} also employs a tunnel diode to achieve a reflective amplifier for communication. However, it cannot be applied to \sysname because it suppresses the harmonics of carriers to reduce interference. The harmonics suppression would lead to distortion of GPS signals since GPS signals are spectrum-spread.

\vspace{-0.3cm}
\section{Discussion} \label{sec: Discussion} 

\textbf{Positioning error.}
\sysname demonstrates significant improvement in localization accuracy compared to other GPS solutions, however, there is still room for further improvement to reduce positioning errors. The source of the positioning error of \sysname may include (i) the average user range error (URE) set by government limitations, which is approximately 0.6m~\cite{userRangeError}, (ii) wrong measurements due to low SNR, (iii) insufficient spatial diversity with the use of a single tag. The limitations (i) and (ii) are fundamental to GPS signals, while (iii) can be addressed through the deployment of multiple tags. In addition, the accuracy can be further improved by applying continuous-time Kalman filters and sensor-fusion techniques.

\textbf{Deployment cost.}
The \sysname has several advantages over traditional GPS relays/repeaters in terms of cost efficiency for deployment. First, \sysname does not need a plug-in power supply, which can significantly reduce system deployment and maintenance costs. Second, \sysname costs $\$$55.16 and 82$\%$ of the tag’s cost comes from tunnel diodes, which will further “reduce cost if developed at scale”~\cite{varshney2020tunnel}, while a commercial GPS relay/repeater system with similar coverage under regulations will cost more~\cite{GPSrelayGRK, GPSrelayRGA30-DV}. 
Besides, \sysname also has a lower deployment cost than BLE beacons in GPS-shadowed regions. Typically, only one \sysname tag is needed to provide positioning services for smartphones, whereas at least three BLE beacons are required for the same purpose. The cost of a \sysname tag is cheaper than BLE beacons (such as an iBeacon, which costs $\$$69~\cite{IBeacon} for a set of 3). Therefore, \sysname is more cost-efficient for deployment.

\vspace{-0.2cm}
\section{Conclusion}
We designed and validated \sysname, an ultra-low power system to expand accurate GPS positioning to shadowed and indoor regions for smartphones. The key enabling hardware in \sysname is the first high-sensitive backscatter that can re-radiate extremely weak GPS signals and provide enough coverage that approaches the regulation limit. 
Beyond GPS positioning, we also envision the \sysname design can be adapted to accommodate a wider range of use cases, such as satellite communications through backscatter.

\vspace{-0.1cm}
\section*{Acknowledgments}

We thank our shepherd and reviewers for their helpful feedback, which greatly improved the quality of the paper.
This work was supported in part by the National Key R$\&$D Program of China under Grant 2020YFB1806600, the National Science Foundation of China with Grant 62071194, the Key R$\&$D Program of Hubei Province of China under Grant No. 2021EHB002, the Knowledge Innovation Program of Wuhan-Shuguang, the European Union's Horizon 2020 research and innovation programme under the Marie Skłodowska-Curie grant agreement No 101022280 and No 824019.

\bibliographystyle{ACM-Reference-Format}
\bibliography{reference}

\end{document}